\documentclass[usegraphicx,usenatbib]{mn2e} \voffset -0.7truecm
  at
10truept

\title [Triaxial haloes, intrinsic alignments and the dark matter
power spectrum ] {\vglue -3.0truecm \centerline{ }\vglue 2.5truecm
Triaxial haloes, intrinsic alignments and the dark matter power
spectrum}

\author
[{\it R.~E.~Smith \& P.~I.~R.~Watts}]
{\parbox[t]{\textwidth}{
    R.~E.~Smith$^{1,3,4}$ \thanks{res@astro.physics.upenn.edu} 
    \& P.~I.~R.~Watts$^{1,2}$ \thanks{pwatts@astro.uni-bonn.de}\\
    {\small 
      $^1$ School of Physics and Astronomy,
      University of Nottingham,
      University Park,
      Nottingham, NG7 2RD, UK.\vspace{-0.2cm}\\
      $^2$ Institut F\"ur Astrophysik und Extraterrestrische Forschung,
      Universit\"at Bonn, Auf dem H\"ugel 71,
      D-53121, Bonn, Germany.\vspace{-0.2cm}\\
      $^3$ Department of Physics and Astronomy,
      University of Pennsylyvania,
      209 South 33rd Street, Philadelphia, PA 19104, USA. \vspace{-0.2cm} \\
      $^4$ Physics and Astronomy Department,
      University of Pittsburgh, 100 Allen Hall, 3941 O'Hara Street,
      Pittsburgh, PA 15260, USA.
    }
  }
}

\def\bk{{\bf k}}
\def\br{{\bf r}}
\def\bx{{\bf x}}
\def\c{\rm c}
\def\s{\rm s}
\def\dirac{{\delta^D}}
\def\dcol{\delta_{\rm col}}
\def\abc2{{\left(\frac{ab}{c^2}\right)}}

\def\a{{\bf a}}
\def\d3x{{d^3\!x}}

\def\be{\begin{equation}}
\def\ee{\end{equation}}
\def\rhob{\bar{\rho}}
\def\rhoc{\rho_{\rm crit}}
\def\nbar{\bar{n}}
\def\simless{\mathbin{\lower 3pt\hbox
  {$\rlap{\raise 5pt\hbox{$\char'074$}}\mathchar"7218$}}}   
\def\simgreat{\mathbin{\lower 3pt\hbox
   {$\rlap{\raise 5pt\hbox{$\char'076$}}\mathchar"7218$}}}  

\def\ba{\begin{eqnarray}}
\def\ea{\end{eqnarray}}
\def\NFW{\rm NFW}
\def\JS{\rm JS}
\def\ST{\rm ST}
\def\vir{\rm vir}
\def\bess0{{\rm J_0}}
\def\sbess0{{\rm j_0}}

\newcommand{\avec}{\mbox{\boldmath $\mathcal{E}$}}

\newcommand{\compi}{{\mathrm i}}

\begin{document}

\maketitle


\begin{abstract}\vspace{-0.3cm}\\
  We develop the halo model of large-scale structure to include
  triaxial dark matter haloes and their intrinsic alignments. As a
  direct application we derive general expressions for the two-point
  correlation function and the power spectrum. We then focus on the
  power spectrum and numerically solve the general expressions for two
  different models of the triaxial profiles.  The first is a toy-model
  that allows us to isolate the dependence of clustering on halo shape
  alone and the second is the more realistic profile model of
  \citet{JingSuto2002}. In both cases, we find that the effect of
  triaxiality is manifest as a suppression of power at the level of
  $\sim5\%$ on scales $k\sim 1-10\,h {\rm Mpc}^{-1}$, which in real
  space corresponds to the virial radii of clusters.  When considered
  by mass, we find that for the first model the effects are again
  apparent as a suppresion of power and that they are more significant
  for the high mass haloes. For the Jing \& Suto model, we find a
  suppression of power on large scales followed by a sharp
  amplification on small scales at the level of $\sim10-15\%$.
  Interstingly, when averaged over the entire mass function this
  amplification effect is surpressed.  We also find for the 1-Halo
  term on scales $k<10\,h {\rm Mpc}^{-1}$, that the power is dominated
  by ellipsoidal haloes with semi-minor to semi-major axis ratios
  $a/c<0.7$.
  
  One of the important features of our formalism is that it allows for
  the self-consistent inclusion of the intrinsic alignments of haloes.
  The alignments are specified through the correlation function of
  halo seeds.  We develop a useful toy model for this and then make
  estimates of the alignment contribution to the power spectrum.
  Further, through consideration of the (artificial) case where all
  haloes are perfectly aligned, we calculate the maximum possible
  contribution to the clustering. We find the hard limit of $<10\%$.
  Subject to further scrutiny, the proposed toy-model may serve as a
  means for linking the actual observed intrinsic alignments of
  galaxies to physical quantities of interest.
\end{abstract}



\section{Introduction}

It is well known that the correlation functions for the dark matter
contain a wealth of information about the cosmological model, the
physics of the dark matter and the relative mix of baryons, photons
and neutrinos
\citep[see][]{BondEfstathiou1984,SeljakZaldarriaga1996,EisensteinHu1998}.
However, it is less well known that they also contain information
concerning the shapes of the structures that form. Under the
assumption that the initial density field is Gaussian random, then
during the linear stages of gravitational collapse all of the
correlation functions, except for the 2-pt function, are zero and the
fluctuations are effectively shapeless.  However, under continued
gravitational collapse, asymmetries develop and are amplified. For the
hierarchical Cold Dark Matter (CDM) models this results in halo
formation and then the emergence of the `cosmic web'
\citep{BondMyers1996}. Subsequently, all of the higher order
correlation functions are non-zero and must now contain information
about the morphological structure of the density field.

Recent attempts to model the 2-pt correlation function have shown that
qualitatively this can be matched with no further assumption than:
that the haloes that form are spherical with some particular density
profile \citep[hereafter NFW]{NavarroFrenkWhite1997}; that the
large-scale over-densities are ellipsoidal \citep{ShethMoTormen2000};
and that halo formation is preferential in the regions of large-scale
over-density \citep[hereafter ST]{ShethTormen1999}.  However, when
this approach is used to model the 3-pt correlation function,
significant discrepancies are found \citep{Scoccimarroetal2001}.
Currently, the root of these discrepancies is believed to be simply
due to the triaxiality of dark matter haloes. However, this has yet to
be explicitly shown.

Furthermore, whilst the 2-pt statistics can be qualitatively
reproduced with simplified models, it still remains to be seen whether
these approaches can be made to precisely match results in general
\citep{Smithetal2003,HuffenbergerSeljak2003}.  With current galaxy
redshift surveys, such as the 2-degree Field Redshift Survey
\citep{Collessetal2001} and the Sloan Digital Sky Survey
\citep{Straussetal2002}, being sufficiently large enough to produce
hi-fidelity measurements of the 2-pt
\citep{Percivaletal2001,Hawkinsetal2003,Zehavietal2002,Tegmarketal2004}
and 3-pt galaxy clustering statistics
\citep{JingBorner2004,Kayoetal2004}, and with planned cosmic weak
shear surveys expected to be capable of measuring the projected 2-pt
and 3-pt matter clustering statistics to similar levels of accuracy
\citep{Alderingetal2004}, resolving these issues becomes important.

In this and subsequent work, we will explore these questions in
detail.  Our first aim, in this paper, is to develop a self-consistent
analytic model that allows us to make predictions for the nonlinear
dark matter clustering signal, given information about the
cosmological model and now, additionally, the shapes of dark matter
haloes and also their intrinsic alignments. A second aim will be to
provide, as an application of the model, predictions for the 2-pt
clustering statistics.  We achieve these goals by developing the
halo-model of large-scale structure
(\citealt{Seljak2000,PeacockSmith2000,MaFry2000}; and for a review see
\citealt{CooraySheth2003}).

An important by-product of this work is that the inclusion of halo
shapes allows one to study the intrinsic alignments of dark matter
haloes alone. This is achieved through the inclusion of the
correlation function of halo axis direction vectors, and we develop a
toy-model to explore the effects of halo alignment on the power
spectrum. 

The outline of this paper is as follows. In Section
\ref{sec:definitions}, we define some useful theoretical notions.  In
Section \ref{sec:formalism} we present the triaxial halo model
formalism and provide a derivation for the 2-pt correlation function.
In Sections \ref{sec:powerspec}--\ref{sec:2halo} we derive the power
spectrum. In Section \ref{sec:details}, we flesh-out the necessary
details for performing calculations. In particular we consider two
interesting models for the density profile of triaxial haloes: the
first is a toy model that we have developed to explore how halo shape
alone affects the clustering statistics; the second is the more
realistic model of \citet[hereafter JS02]{JingSuto2002}. Here, we also
develop our toy model for the intrinsic alignments correlation
function. In Section \ref{sec:results}, we present our results, and
finally in Section \ref{sec:discussion} we discuss our findings and
present the conclusions. 

Throughout, we have assumed that the cosmological model is the
concordance model \citep{Wangetal2000} and that the linear power
spectrum is given by \citet{Efstathiouetal1992} with normalization
$\sigma_8=0.9$ and shape parameter $\Gamma=0.21$.


\section{Theory}

\subsection{Basic definitions}\label{sec:definitions}

In what follows, we will seek to compute the lowest order clustering
statistic of interest, that is the power spectrum of mass
fluctuations, $P({\bf k})$. This is defined by
\be \langle \delta(\bk)\,\delta(\bk')\rangle = (2\pi)^3P({\bf k})\,
\delta^D(\bk - \bk') \ ,\label{powerdef}\ \ee
where $\delta(\bk)$ is the Fourier transform of the density
fluctuation field $\delta(\bx)=\rho(\bx)/\bar{\rho} - 1$ and $\rhob$
is the background density; angle brackets denote the ensemble average
and $\delta^D$ is the Dirac delta function.  The Power Spectrum is
itself the Fourier transform of the real space 2-pt correlation
function $\xi({\bf r})$, which can be similarly defined
\be \xi({\bf r}) = \langle \delta(\bx)\,\delta(\bx + \br) \rangle
\label{correldef}\ .
\ee
In the above definitions we have considered general anisotropic
density fields. However, in cosmological applications it is usual to
assume statistical isotropy and homogeneity of the Universe, hence
these quantities become functions of scalar arguments only. In the
work that follows, we will not make the above assumptions, since we
are dealing with anisotropic dark matter haloes, but will show that
for the power spectrum the scalar arguments that we desire will
naturally emerge from our formalism.

It will also prove useful to define some basic relations for the
triaxial haloes \citep[see][for a full treatise]{Chandrasekhar1969}.
We start by considering an heterogeneous triaxial ellipsoid that has
semi-axis lengths $a$, $b$ and $c$, where $a \leq b \leq c$, and
orthogonal principle axis vectors $\hat{\bf{e}}_a$, $\hat{\bf{e}}_b$
and $\hat{\bf{e}}_c$.  We define the triaxial coordinate system
$(R,\Theta,\Phi)$ with respect to the principle axes of the halo,
where $\hat{\bf{e}}_c$ is taken to be in the $z$-direction.  The
radial parameter $R$ traces out thin iso-density shells, or homoeoids,
and the parameters $\Theta$ and $\Phi$ are the polar and azimuthal
angles respectively. In this system the Cartesian components are
\be x = \frac{a}{c} R \cos \Phi \sin \Theta \ ; \ y = \frac{b}{c} R
\sin \Phi \sin \Theta \ ; \ z = R \cos\Theta.
\label{ellipdef}\ee
It is to be noted that the ellipsoidal angles differ from those of the
spherical coordinate system. The parameter $R$ can be related to the
Cartesian coordinates and axis ratios through
\be \frac{R^2}{c^2} = \frac{x^2}{a^2} + \frac{y^2}{b^2} + \frac{z^2}{c^2}
\label{isodef}\ .\ee
The benefits of this choice of coordinate system are now apparent: if
all the homoeoidal shells are concentric, and if we pick coordinates
with the same axis ratios as the triaxial ellipsoid, then the density
run of the ellipsoid can be described by a single parameter:
\be \rho(\br)\rightarrow\rho(R)\ .\label{eq:rexchange}\ee
Furthermore, the mass enclosed within some iso-density cut-off scale
$R_{\rm cut}$, can be obtained most simply by
\be M=\int_{R_{\rm cut}(\br)} d^3\!r \rho(\br)=4\pi \frac{ab}{c^2}
\int_0^{R_{\rm cut}} dR R^2 \rho(R) \ ,\ee
where the ellipsoidal coordinates have allowed us to circumvent the
problem of evaluating complicated halo boundaries.

It is now convenient to define what we mean by a halo: any object
that has a volume averaged over-density 200 times the background
density is considered to be a gravitationally bound halo of dark
matter. This leads directly to the following relation between the
mass, radius and axis ratios:
\be M_{200}=\frac{4}{3}\pi R_{200}^3 \abc2 200 \rhob \ 
.\label{eq:M200}\ee
The above definition was adopted in order to be consistent with the
mass-function of ST, which we utilize in what follows.


\subsection{The triaxial halo model}\label{sec:formalism}

In earlier implementations of the halo-model it was assumed that all
of the matter in the Universe was contained within spherical dark
matter haloes with some particular density profile and some
distribution of mass \citep{Seljak2000,PeacockSmith2000,MaFry2000}. We
now re-develop the halo model formalism making the important
modification that haloes are not in general spherical, but instead are
more closely described by the family of triaxial ellipsoids.

To start, we characterize each halo in terms of a set of stochastic
variables that describe both shape and orientation: a given halo of
mass $M$ will therefore have principle axis vectors $\hat{\bf{e}}_a$,
$\hat{\bf{e}}_b$ and $\hat{\bf{e}}_c$ with axis ratios $a/c$ and
$b/c$. The density run of a particular halo is described by the
function $\rho(\br,M,\avec,\a)\equiv M\,U(\br,M,\avec,\a)$, where $U$
is the mass normalized profile, and we have introduced the shorthand
notation $ \avec \equiv ( \hat{\bf{e}}_a , \hat{\bf{e}}_b ,
\hat{\bf{e}}_c )$ and $\a \equiv (a,b,c)$. The density at any point
can now be expressed simply as a sum over the $N$ haloes that form the
field
\be \rho(\br) = \sum_i^N M_i\, U(\br-\bx_i , M_i , \avec_i , \a_i),
\label{sumoverhaloes}
\ee
where ${\bf x_i}$ denotes the position vector of the centre of mass of
the $i^{\rm th}$ halo. Following \citet{ScherrerBertschinger1991}, we
can re-write the sum in equation (\ref{sumoverhaloes}) using the
substitution
\ba \sum_i & \rightarrow & \int d \bx \, d M \, d \avec \, d \a \sum_i
\dirac(\bx-\bx_i)\, \dirac(M-M_i) \nonumber \\ & & \times \,
\dirac(\avec-\avec_i)\, \dirac(\a - \a_i)\ , \label{sumswap} \ea
which allows us to consider integrals over continuous variables rather
than a sum over discrete quantities. The integral over $\avec$ in the
above expression indicates an integral over all possible orientations
of the halo. The orientation of the halo frame can be specified
relative to a fixed Cartesian basis set through the Euler angles.
These represent successive rotations of the halo frame about the
$z$--axis by $\alpha$, the $y'$--axis by $\beta$, and the $z''$--axis
by $\gamma$ \citep[see][]{MathewsWalker1970}. Hence the process of
averaging over all possible halo orientations can be performed by
integrating over all possible Euler angles.

Lastly, to compute the ensemble averages we integrate over the joint
probability density function for the $N$ haloes that form the density
field, and sum over the probabilities for obtaining the $N$ haloes
\citep[see][for a similar approach]{McClellandSilk1977}: %
\[
\langle \cdot\cdot\cdot \rangle \equiv \sum_{j} p(N_j|V) \int
\prod_{i=1}^{N_j} dM_i\;d^3x_i\;d{\bf a}_i\;d\avec_i
\]
\be \hspace{0.5cm} \times p(M_1,..,M_{N_j},{\bf x}_1,..,{\bf x}_{N_j},{\bf
a}_1,..,{\bf a}_{N_j}, \avec_1,..,\avec_{N_j}|N_j)\ . \label{eq:ensemble}\ee
Provided the volume of space considered is large, then $p(N|V)$ is very
sharply spiked around $N=\nbar V\gg 1$, where $\nbar$ is the mean number
density of haloes. We restrict our study to this case only. Also, the
ensemble average of equation (\ref{sumoverhaloes}) gives the mean
density of the universe, $\bar{\rho}$.

To compute the 2-point correlation function of the triaxial halo field
we substitute equations (\ref{sumoverhaloes}) and (\ref{sumswap}) into
(\ref{correldef}) and compute the ensemble average according to
equation (\ref{eq:ensemble}).  The result is a sum of two terms, the
first accounts for the correlation between points in a single halo and
the second accounts for the inter-halo correlation (hereafter the
1-Halo and 2-Halo terms):
\be \xi({\bf r}) = \xi^{\rm 1H}({\bf r}) + \xi^{\rm 2H}({\bf r}) \ :
\label{correlfull}\ee
\ba \xi^{\rm 1H}({\bf r})\!\!\!\! & = \!\!\!\! &
\frac{N}{\bar{\rho}^2} \int d M \, d\bx \, d \avec \, d \a \; M^2\,
p(\bx, M, \avec, \a) \nonumber \\
& & \times \; U(\br_1 - \bx, M, \avec, \a) \;
U(\br_2 - \bx, M, \avec, \a) -1 \ ;\label{correl1halmost}\ea
\ba \xi^{\rm 2H}({\bf r})\!\!\!\! & = \!\!\!\! &
\frac{N^2}{\bar{\rho}^2} \int d M_1 \, d M_2 \, d\bx_1 \, d\bx_2 \, d
\avec_1 \, d \avec_2 \, d \a_1 \, d \a_2 \,M_1\, \,M_2\, \nonumber \\
& & \times \; U(\br_1 - \bx_1, M_1, \avec_1, \a_1)
\,U(\br_2 - \bx_2, M_2, \avec_2, \a_2) \nonumber \\
& & \times \; p(\bx_1, \bx_2, M_1, M_2, \avec_1,
\avec_2, \a_1, \a_2)-1\ , \label{correl2halmost}\ea
where the separation vector $\br = \br_1 - \br_2 $.

The joint probability density function in the 2-Halo term may be
written:
\be p(1,2)=p(1)\,p(2)\left[1+\xi^{\s}(1,2)\right] \ ,\label{jointprob}\ee
%
where we have adopted the short-hand notation $ p(1)\equiv p(\bx_1,
M_1, \avec_1, \a_1)$, and where $\xi^{\rm s}$ is the seed-correlation
function, which describes the relationship between a particular halo's
characteristics and those of all the other haloes. In the spherical
halo model this would just be the halo-bias of \cite{MoWhite1996},
however in the triaxial model the function is more complicated, with
halo orientation vectors and shapes being influenced by those of
neighbouring objects. We explore this in greater detail in Section
\ref{sec:2halo}.

Neglecting for the time being $\xi^{\s}(1,2)$, all that remains to
arrive at the correlation function in the triaxial halo model is
to deal with the joint probability density function for a single
halo's characteristics: $p(1)$. We assume that a halo's
orientation, position and mass are independent random variables,
and that the halo axis ratios are dependent on mass only. Hence we
have:
\be  p(\bx,M,\avec,\a) =
\frac{1}{V}\frac{n(M)}{\nbar}p(\avec)p(\a|M) , \label{probone} \ee
where $n(M)$ is the halo mass function. In order to provide a uniform
probability for the halo orientation on the sphere, the density
function $p(\avec)$ takes the form
\ba p(\avec)d\avec & \equiv & p(\alpha,\beta,\gamma)d \alpha\, d
\beta\,d \gamma \ ,\nonumber \\
& = & \frac{1}{2\pi} \frac{1}{2}\frac{1}{2\pi}\, d \alpha\, d (\cos\beta)
\,d \gamma, \label{probavec}\ea
where the variables are restricted to the ranges: $0\ge \alpha \ge
2\pi$, $0\ge\beta\ge\pi$ and $0\ge \gamma \ge 2\pi$. On substituting
equations (\ref{jointprob}) and (\ref{probone}) into
(\ref{correl1halmost}) and (\ref{correl2halmost}) we find:
\ba \xi^{\rm 1H}({\bf r})\!\! & = & \frac{1}{\bar{\rho}^2} \int d M \,
d\bx \, d \avec \, d \a \, M^2\, n(M)\, p(\avec)\, p(\a|M)\nonumber \\
& & \times \,U(\br_1 -\bx) U(\br_2 - \bx)
\label{correl1hresult} \ ;\ea
\ba \xi^{\rm 2H}({\bf r}) & = & \frac{1}{\bar{\rho}^2} \int d M_1 \,
d M_2 \, d\bx_1 \, d\bx_2 \, d \avec_1 \, d \avec_2 \, d \a_1 \, d
\a_2 \, M_1\,M_2\, \nonumber \\
& & \times \;  n(M_1) \,n(M_2)\, p(\avec_1) \,
p(\avec_2)\, p(\a_1 | M_1)\, p(\a_2 | M_2)\, \nonumber \\
& & \times \xi^{\rm s}(1,2)\,U_1(\br_1 - \bx_1) \,U_2(\br_2 - \bx_2)\
;\label{correl2hresult} \ea
where we have suppressed the explicit functional dependence of the
density profile $U$ on all variables except the position vector and
the subscripts refer to either halo one or two.


\subsection{The dark matter power spectrum}\label{sec:powerspec}

The power spectrum of the triaxial halo field can be evaluated by
taking the Fourier transform of expressions (\ref{correl1hresult})
and (\ref{correl2hresult}) to yield:
\be
P({\bf k}) = P^{\rm 1H}({\bf k}) + P^{\rm 2H}({\bf k})\ ;\label{powerfull}
\ee
\ba
P^{\rm 1H}({\bf k}) & = & \frac{1}{\bar{\rho}^2} \int d M \, d \avec
\, d \a \; M^2 n(M) \, p(\avec)\, p(\a|M) \nonumber \\
& & \times \tilde{U}(\bk) \, \tilde{U}^*(\bk) \ ;\label{power1hresult}
\ea
\ba P^{\rm 2H}({\bf k}) & = & \frac{1}{\bar{\rho}^2} \int d M_1 \, d
M_2 \, d \avec_1 \, d \avec_2 \, d \a_1 \, d \a_2 \, M_1 \, M_2\, \nonumber \\
& & \times \;  n(M_1) \, n(M_2) \, p(\avec_1)\,
p(\avec_2)\, p(\a_1 | M_1) \, p(\a_2 | M_2) \,
\nonumber \\
& & \times \; P^{\rm s} (1,2) \,\tilde{U}(\bk) \, 
\tilde{U}^*(\bk), \label{power2hresult}
\ea
where $\tilde{U}$ is the Fourier transform of the halo profile and
$P^{\rm s}(1,2)$ is the Fourier transform of $\xi^{\rm s} (1,2)$.


\subsection{The 1--Halo term}\label{sec:1Halo}

In this Section we focus on the 1-Halo term of equation
$(\ref{power1hresult})$. In order to solve this equation in its
present form we are required to perform a 12-D numerical integration.
Evaluation of the complete expression would therefore be both
numerically noisy and slow. However, we can use properties of the
elliptical coordinate system defined in Section \ref{sec:definitions}
to simplify the general result significantly.

We begin by re-writing the 1-Halo term of equation
(\ref{power1hresult}) in the form
\be P^{\rm 1H}({\bf k}) = \frac{1}{\bar{\rho}^2} \int d M d \a \, M^2
n(M) \, p(\a|M)\, W({\bf k})\ , \ee
where the window function is defined to be 
\be W({\bf k}) = \int d\avec \,p(\avec) \, \left| \int d\br \;
U(\br,\avec) \exp{\compi\bk\cdot\br} \right|^2.\ee
We now observe the following symmetry: averaging over all halo
orientations for a fixed vector $\bk$ is equivalent to averaging over
the direction vector $\hat{\bk}$ at fixed halo orientation.  Hence, we
are able to replace the integral $\int d\avec p(\avec)\rightarrow \int
d \hat{\bk}/4\pi$. This means that $W$ and hence $P^{\rm 1H}$ are now
simply functions of the scalar length $k$.  Consider next the $\br$
integral; if we transform to the ellipsoidal coordinates and recall
from earlier that under such a transformation the halo profile is
simply a function of iso-radius $R$ and also that the iso-density
surface at which the halo is truncated is simply some particular value
of $R=R_{\rm cut}$, then the window function becomes
\be W(k) = \int \frac{d \hat{\bk}}{4\pi} \left| \int \, d R R^2
  U(R) \frac{ab}{c^2} \int d \hat{\bf R}\, \exp \left[ \compi \bk \cdot
  \br({\bf R})\right] \right|^2 \ . \ee
Considering now the $\hat{\bf R}$-integral, if we write the Cartesian
components of the ${\bf r}$ and ${\bf k}$ vectors in terms of
ellipsoidal and spherical coordinates as
\be \br = R \left( \frac{a}{c} \cos \Phi \sin \Theta , \frac{b}{c}
\sin \Phi \sin \Theta , \cos \Theta\right)\ ,\ee
\be \bk = k(\cos\phi_k \sin\theta_k , \sin \phi_k \sin \theta_k , \cos
\theta_k )\ ,\ee
then the integrals may be evaluated using the standard results
\citep{GradshteynRyzhik1994}:
\be \int_0^{2\pi} d\phi \, \exp(\compi u \cos \phi+\compi v \sin\phi) 
 = 2\pi \bess0\!\left(\sqrt{u^2 + v^2}\, \right) \ ;\ee
\be \int_0^1 d x \cos (u x) \bess0\!\left(v \sqrt{1-x^2}\, \right) =
\sbess0\left(\sqrt{u^2+v^2}\, \right),
\ee
where ${\rm J_0}$ is a Bessel function of the first kind and ${\rm
  j_0}$ is a spherical Bessel function.  After a little algebra, we
arrive at the final result for the 1-Halo component of the power
spectrum:
\be P^{\rm 1H}(k) = \frac{1}{\bar{\rho}^2} \int\,d M \, d \a \, n(M)\,
 M^2 \, p(\a|M) W(k) \ ,\label{power1H}\ \ee
where 
\be W(k)=\int \frac{d\hat{\bk}}{4\pi} \left[ \int d R R^2 \,
\frac{ab}{c^2} U(R) \, 4\pi \sbess0\left[k R f(\theta_k,
\phi_k)\right] \right]^2 \label{windowfunction} \ee
and where
\be f^2(\theta_k, \phi_k) = \cos^2 \theta_k + \sin^2 \theta_k \left(
  \frac{a^2}{c^2}\cos^2 \phi_k + \frac{b^2}{c^2} \sin^2\phi_k \right).
\label{fthetaphi} \ee
It is informative to consider two limits of the above formula.
Firstly, if haloes are spherical, $a/c=b/c=1$, then we recover the
standard expression for the 1-Halo term \citep{PeacockSmith2000}.
Secondly, if we consider the limit $kR \ll 1$, then the $\sbess0$-term is
approximately unity and we have, as required, the power simply being
related to the effective number density of objects
\citep{PeacockSmith2000}.

Through these efforts we have thus reduced the dimensionality of our
integral from 12-D to 6-D, and in Section \ref{sec:results} we
recommend two numerical methods for solving integrals of this type.


\subsection{The 2--Halo term}\label{sec:2halo}

We now turn our attention to the 2-Halo term. Again, our aim is to
manipulate the general result of equation (\ref{power2hresult}) into a
form that is more amenable to direct computation. However, we now have
to deal with the added complication of possible intrinsic correlations
in halo properties and this will require us to provide a detailed form
for $\xi^{\rm s}(1,2)$.


\subsubsection{No alignments}

To begin with it will prove useful to consider the simplest possible
case, namely that for which there are \emph{no} correlations between
the halo seeds other than those between position and mass. In this
instance the seed correlation function reduces to that of the
spherical halo model and we may write
\citep[ST]{MoWhite1996},
\be P^{\s}({\bf k},M_1,M_2,\avec_1,\avec_2,\a_1,\a_2) = b(M_1)\,
b(M_2)\, P^{\rm L}(k)\ ,\label{halobias}\ee
where $P^{\rm L}$ is the linear dark matter power spectrum, and $b(M)$
is the bias function for haloes of mass $M$. On inserting this into
equation (\ref{power2hresult}), and once again using the symmetry
between averages over $\avec$ and $\hat{\bf k}$, we find that the
2-Halo term with no intrinsic alignments can be written
\be P^{\rm 2H}_{\rm NA}(k)= \left[ \frac{1}{\rhob} \int d M\, M n(M)\,
b(M) \,\tilde{\mathcal U}(k,M)\right]^2 P^{\rm L}(k)
\label{power2hnocorr}\ee
where 
\be \tilde{\mathcal U}(k,M)=\frac{1}{4\pi}\int d \a\, p(\a|M) \int d
\hat{\bf k} \,U({\bf k},M,\a,\avec) \ .\label{SphereAveragProfile}\ee
Equation (\ref{power2hnocorr}) is identical to the standard spherical
halo model result \citep{Seljak2000}. Thus we may understand the
2-Halo term, for this case, to be simply the correlation of the
spherically averaged triaxial haloes. However, owing to the
complication of determining the spherically averaged profiles and the
associated virial radius, we provide an alternate expression for the
2-Halo term that yields more easily to direct computation. Considering
again equation (\ref{power2hresult}), if we take the inverse Fourier
transforms of the density profiles, transform from the spherical polar
coordinates to the ellipsoidal, and interchange the integral over the
halo Euler angles to an average over $\hat{\bf k}$, then we find
\ba P^{\rm 2H}_{\rm NA}(k) \! & = \! & \!\ P^{\rm
L}(k)\left[\frac{1}{\rhob}\int dM M n(M) \, b(M) \int d\a p(\a|M)
\right. \nonumber \\ & & \hspace{-15mm} \left. \times \, \int dR
R^2 \left(\frac{ab}{c^2}\right) U(R)\int d\hat{\bf k}\,
\sbess0\left[k R f(\theta_k, \phi_k)\right]\right]^2
\label{power2Hnoalign}\ .\ea


\subsubsection{With intrinsic alignments}\label{ssec:align}

Although the alignment of galactic/halo spins is an old subject,
dating back to work by \citet{Hoyle1949}, it has received relatively
little attention over the past fifty years. However, in recent times
it has been the focus of much work. This renewed interest has mainly
been driven by a need to understand the possible contamination that
galaxy shape alignments may contribute to cosmic weak shear surveys
\citep{Brownetal2002,Heymansetal2004}. Despite this renewed interest,
the detailed physics underlying the process still remains the subject
of much debate \citep{LeePen2001,Mackeyetal2002,Porcianietal2002}, and
as such no universally accepted model has yet been identified
\citep{Heymansetal2004}. Moreover, these proposed models are not
specified in a form that allows for direct insertion into our
formalism.  Alternatively, a number of authors have explored the
alignment problem through direct numerical simulation.
\citet{CroftMetzler2000,Heavensetal2000,Jing2002} have explored the
correlation of halo ellipticities projected onto the sky and
\citet{HattonNinin2001,Hopkinsetal2004} have studied the correlation
of the scalar product of the semi-major axis vectors with separation.
However, no simple phenomenological model for the alignment
correlation function, in the form that we require, exists. Thus in
order to proceed our strategy will be to develop a toy-model for the
intrinsic alignment correlation function $C(|\hat{\bf e}_c^1\cdot
\hat{\bf e}_c^2|)$ that will allow us to probe, in a qualitative
sense, the effects of halo alignments on clustering.

We now outline some of the assumptions upon which the model is based
since these will be useful in what follows, but we reserve the
complete details to Section \ref{sec:details}.  Firstly, we assume
that the intrinsic alignment between two haloes is simply manifest as
a correlation of the semi-major axis vectors ${\bf \hat{e}_c}$; the
${\bf \hat{e}_b}$ and ${\bf \hat{e}_a}$ axes we will assume are
uniformly random. Secondly, we assume that the spatial correlation
function of the halo centres is independent of the halo orientations.
Thirdly, we assume that the axis ratios, $\a$, are uncorrelated from
one halo to another, and that these quantities depend only on the mass
of each halo.  Under these assumptions equation (\ref{jointprob})
reduces to
\ba p(1,2) & = & p(1)\,p(2) \left\{1+\xi({\bf x}_1,{\bf
x}_2|M_1,M_2) + \right. \nonumber \\ & & \left.
C(\mu|M_1,M_2)\left[ 1+\xi({\bf x}_1,{\bf
x}_2|M_1,M_2)\right] \right\}\ \label{eq:alignprob},\ea
where $\mu=\hat{\bf e}_c^1\cdot\hat{\bf e}_c^2$.  On inserting the
above expression into equation (\ref{correl2halmost}) for the 2-Halo
correlation function, we find that the first term in the curly
brackets gives the mean density squared, the second gives the 2-Halo
term without alignments, the third integrates to zero through the
integral constraint,
$ \int_{-1}^{1} C(\mu)\,d\mu=0 \label{constraint}$
and the fourth term represents the intrinsic alignment contribution to
the 2-Halo term. This last term can be written,
\ba \xi^{\rm 2H}_{\rm IA}({\bf r})\hspace{-0.2cm} & = \hspace{-0.2cm}&
\frac{1}{\rhob^2}\int dM_1\, dM_2\, d\avec_1 \,d\avec_2\, d \a_1 \,d
\a_2\, d\bx_1\, d\bx_2\, M_1 \,M_2 \nonumber \\
 & & \times \, n(M_1)\, n(M_2) \,p(\a_1|M_1)\,p(\a_2|M_2)\, p(\avec_1)\,
 p(\avec_2)\nonumber \\
& & \times U_1({\bf r}_1-{\bf x}_1) \,U_2({\bf r}_2-{\bf x}_2)
\,\xi^{\s}({\bf r}|M_1,M_2) \nonumber \\
& & \, C(\mu|M_1,M_2) \ .\label{eq:corralign}\ea
Fourier transforming the above expression leads us to the intrinsic
alignment contribution to the power spectrum,
\ba P^{\rm 2H}_{\rm IA}(k) & = & \frac{P^{\rm L}(k)}{\rhob^2} \int
dM_1 dM_2 d \a_1 d\a_2\, M_1 M_2 \, n(M_1)\, n(M_2) \nonumber \\
& & \times b(M_1)\, b(M_2)\, p(\a_1|M)\,p(\a_2|M) F({\bf k})\ ,
\label{powIA}\ea
where, 
\be F({\bf k}) = \int \frac{d\avec_1\, d\avec_2}{(8\pi^2)^2}\,
\,\tilde{U}_1({\bf k},\avec_1)\,\tilde{U}_2({\bf
k},\avec_2)\, C(\mu|M_1,M_2)\ ,\label{fk}\ee
and where we have made use of the halo bias relation given by equation
(\ref{halobias}). Focusing on equation (\ref{fk}), again we can swap
the integral over $\avec_1$ to an average over $\hat{\bf k}$.
However, owing to the alignment correlation function we must now
specify the fixed coordinate system of $\avec_1$, and we let this be
the usual Cartesian system. Next, if we expand $\avec_2$ in terms of
the three Euler angles $\alpha$, $\beta$ and $\gamma$, where these are
now rotation angles relative to the $\avec_1$ system, and on realizing
that $\hat{\bf e}^1_c\cdot \hat{\bf e}^2_c=\mu=\cos\beta$, then we
have that
\be F(k)\! =\! \int \!\frac{d\hat{\bf k}}{4\pi}\frac{d\alpha\, d\mu\,
d\gamma}{8\pi^2} \, \tilde{U}_1({\bf k}) \tilde{U}_2({\bf
k},\alpha,\beta,\gamma)\,C(\mu|M_1,M_2) ,\label{fk2}\ee
where the function 
\ba \tilde{U}_2({\bf k},\alpha,\beta,\gamma) \hspace{-0.25cm} & = &
\hspace{-0.2cm} \frac{ab}{c^2} \int d^3\!R'\, U_2(R') \exp\left[
\compi {{\bf k'}}(\alpha,\beta,\gamma)\cdot {\bf r'}({\bf R'})\right]
\nonumber \\
 & = & \hspace{-0.25cm} \frac{ab}{c^2}\! \int\! dR'\,R'^2 U_2(R') 4\pi
\sbess0\left[kR' f(\theta_k',\phi_k')\right]\!,\label{Utilde}\ea
where ${\bf r}'$ is a vector in the $\avec_2$ frame and where ${\bf
  k}'$ is the $k$-vector rotated from the $\avec_1$ frame into the
$\avec_2$ frame:
\be {\bf k}'={\mathcal R}(\alpha,\beta,\gamma) {\bf k}= {\mathcal
  R}_{z''}(\gamma) {\mathcal R}_{y'}(\beta) {\mathcal R}_z(\alpha)
{\bf k} \ .\ee
${\mathcal R}$ is the $z-y'-z''$ rotation matrix (see Appendix
\ref{app:rotation} for the explicit form) and $\theta_k'$ and
$\phi_k'$ are the spherical polar angles of the $k'$-vector in the
$\avec_2$ frame.  Finally, on inserting equation (\ref{Utilde}) into
(\ref{fk2}) and taking the inverse Fourier transform of $U_1(\bk)$, we
arrive at the final answer for the intrinsic alignments contribution
to the power spectrum
\begin{eqnarray}
P_{\rm IA}^{2\rm H}\!\!\!\!& = \!\!& \!\!\frac{P^{\rm L}(k)}{2\pi
\bar{\rho} ^2} \int dM_1\, dM_2 \, M_1 M_2 \, n(M_1) n(M_2) \, b(M_1)
b(M_2) \nonumber \\ & & \times \int d{\bf a}_1\, d{\bf a}_2 \, P({\bf
a}_1|M_1) P({\bf a}_2|M_2)\, \frac{a_1 b_1}{c^2_1} \frac{a_2
b_2}{c^2_2} \nonumber \\ & & \times \int dR\, dR' \, R^2 R'^2 \,
U_1(R) U_2(R') \nonumber \\ & & \times \int d\hat{{\bf k}} \, {\rm
j_0} [kRf(\theta_k, \phi_k)] \, \mathcal{D}(\theta_k, \phi_k)
\label{power2Halignments}\ ;
\end{eqnarray}
where we have defined the useful function
\be {\mathcal D} = \int d\alpha \, d\mu \, d\gamma \,
\sbess0\left[kR'f(\theta_k',\phi_k')\right] C(\mu|M_1,M_2), \ee
and where for completeness,
\be \theta_k' = \arccos\left[
\frac{\s\beta\left( k_x\c\alpha+k_y\s\alpha\right)+\c\beta k_z}{k}
\right]\ ,\ee
\be \phi_k' = \arccos\left[\frac{\eta}{k \s\theta_k'}\right]\ ,\ee
\be {\rm \eta}=(\c\beta\c\alpha\c\gamma-\s\alpha\s\gamma)k_x
+(\c\beta\s\alpha\c\gamma+\c\alpha\s\gamma)k_y-\s\beta\c\gamma k_z \ \ee
where we adopt the short-hand notation $\s\equiv\sin$ and
$\c\equiv\cos$.

The full power spectrum, incorporating all of the triaxial halo
effects is thus
\be P(k) = P^{\rm 1H}(k) + P^{\rm 2H}_{\rm NA}(k) + P^{\rm 2H}_{\rm
  IA}(k)\ , \ee
where the three terms on the right hand side are given by equations
(\ref{power1H}), (\ref{power2Hnoalign}) and (\ref{power2Halignments})
respectively. These equations together comprise the main results of
this paper. In the next section, we summarize the specific model
details that are required to make direct calculations of these
quantities.


\section{Calculation Details}\label{sec:details}

As mentioned earlier, we use the ST mass function and halo bias
relations. Since these are now widely known, we have reserved all
details to Appendix \ref{app:massfunction}.

\subsection{Models for triaxial density profiles}\label{sec:triaxprofiles}

We consider two models for the triaxial density profiles. The first is
a toy model that we have developed in order to explore how the
clustering statistics are affected by halo shape alone. We refer to
this as the `Continuity' model. The second is the more realistic model
of JS02.

\begin{itemize}
  
\item{\bf The Continuity model:} The main idea here is that triaxial
  density profiles can be generated from the spherical dark matter
  haloes with some specified density profile. To see this, consider
  distorting a spherical halo along two orthogonal axes and let this
  deformation be done in such a way that the volume is preserved. If
  the spherical radial length $r$ is then exchanged for the
  ellipsoidal radial length $R$, given in equation (\ref{isodef}),
  where we use the axis ratios of the ellipsoid formed from the
  deformation of the sphere, then the density structure of the new
  triaxial halo is completely understood in terms of the parameters
  that specified the spherical halo and the axis ratios. In the
  following we will start from the spherical density profile model of
  NFW, 
  \be \rho_s(r)=\frac{\rhoc\;\delta_s^c(M)}{y(1+y)^2}\
  ;\hspace{0.7cm}y\equiv r/r_0(M)\ ,\label{triaxNFW}\ ,\ee 
  where $r_0$ is the scale radius and $\delta_s^c$ is the
  characteristic density contrast (hereafter subscripts $e$ and $s$
  represent the ellipsoidal and spherical models, respectively).  If
  we distort this following the recipe described above then we get the
  ellipsoidal halo profile
  \be \rho_e(R)=\frac{\rhoc\;\delta_e^c(M)}{y(1+y)^2}\
  ;\hspace{0.7cm}y\equiv R/R_0(M)\ ,\label{triaxNFW}\ee
  where $R_0(M)$ and $\delta_e^c(M)$ are the ellipsoidal scale radius
  and characteristic density contrast.
%
%
  As is the case for the spherical profile, $R_0(M)$ and
  $\delta_e^c(M)$ are not independent, but are related through the
  continuity of mass.  Hence,
  \be \delta_e^c=\frac{200 c_e^3/3}{\log(1+c_e)-c_e/(1+c_e)}\
  \label{eq:charden},\ee
  where $c_e\equiv R_{200}/R_0$ is the concentration parameter. One is
  then left with determining a single free parameter. We now realize
  that if the volume is preserved under the distortion, then the
  physical density at the scale radius must be equivalent for the
  spherical and ellipsoidal haloes, hence $\rho_s(r_0)=\rho_e(R_0)$,
  which implies that $\delta_e^c(M)=\delta_s^c(M)$. We therefore
  simply require a model for $\delta_s^c$ to specify the triaxial
  haloes.  NFW found this was well fit by
  \be \delta_{s}^{c}=3000(1+z_c)^3 \ ,\ee
  where $z_c$ is the redshift of collapse, which can be determined
  from the arguments of \cite{LaceyCole1993}.  Finally, the cut-off
  radius of the ellipsoidal haloes can be easily obtained from
  continuity of the mass of spherical and triaxial haloes and is given
  by
  \be R_{200}=\left(\frac{ab}{c^2}\right)^{-1/3}r_{200} 
  \label{triaxviralrad}.\ee

  This model would be complete if the masses of ST and NFW haloes were
  equivalent. However they are not. The NFW halo mass is defined to be
  where the volume averaged overdensity reaches $200\rhoc$, as opposed
  to 200$\rhob$ for the ST case. If one considers a halo that has a
  density run given by the NFW model, then the different mass
  definitions simply correspond to two different truncation
  densitys. Thus one may determine the mapping between the two mass
  definitions through using the NFW profile and again equating the
  physical density at the scale radius. This time the importance of
  using the physical density is that the density contrast relation is
  now modified by a factor of $\Omega$.  Hence one finds
  \be \delta_{\ST}^c=\delta_{\NFW}^c/\Omega\ .\ee
  Using equation (\ref{eq:charden}) we find the relation
  \be \frac{1}{\Omega}\left(\frac{c_{\NFW}}{c_{\ST}}\right)^3
  \left[\frac{\log(1+c_{\ST})-c_{\ST}/(1+c_{\ST})}
  {\log(1+c_{\NFW})-c_{\NFW}/(1+c_{\NFW})}\right]=1\ ,\ee
  where $c_{\NFW}=r^{\rm NFW}_{200}/r_0$ and where $c_{\ST}=r^{\rm
  ST}_{200}/r_0$. This can be solved numerically to give the mapping
  between the two different concentration parameters and thus the
  mapping between the masses. In practice, once we have established
  the mass transformation we fit a spline function to it and use the
  relation
  \be c_{\ST}=\left(\frac{1}{\Omega}\frac{M_{\ST}}{M_{\NFW}}
  \right)^{1/3}c_{\NFW}\ \ee
  to convert between the different concentration parameters.

  Lastly, in this model we do not specify the probability
  distributions for the axis ratios.  Instead, we simply take these
  from the prescription of JS02, which is described below.

  \hspace{\parskip}


\item{\bf The JS02 model:} 
  A more realistic model for triaxial dark matter haloes was presented
  by JS02, who fitted an ellipsoidal NFW density profile to haloes
  measured directly from high resolution numerical simulations.
  Details of their model can be found in the original paper by JS02,
  but are nicely summarized in \cite{OguriLeeSuto2003}.  We discuss
  the points that are relevant for our application and refer the
  reader to these primary sources for a more complete account.
  
  The density run in the JS02 model has the same general form as the
  ellipsoidal NFW profile given in equation (\ref{triaxNFW}).
  However, now the normalization parameters $\delta_e^c(M)$ and
  $R_0(M)$ are independent quantities and are to be determined as
  follows. Firstly, the virial radius of a spherical halo with
  equivalent mass to the triaxial halo is found. From this a scale
  length $R_e$ is determined. This is defined to be a fixed fraction
  of the spherical halo's virial radius,
  \be R_e=0.45 r_{\vir}\label{ellipsoidalR}\ .\ee
  JS02 state that this corresponds to a scale where the volume
  averaged over-density within $R_e$ is given by
  \be
  \Delta_e=5\Delta_{\vir}(\Omega)\left(\frac{ab}{c^2}\right)^{0.75}\ 
  ,\ee
  where $\Delta_{\vir}(\Omega)$ is the over-density of virialization
  from the spherical collapse model. The characteristic density is
  then found through
  \be \delta_e^c=\frac{\Delta_e \, c_e^{3}/3}{\log(1+c_e)-c_e/(1+c_e)}
  \ ;\ \ \ c_e\equiv R_e/R_0(M)\label{eq:deltaedef}\ee
  where $c_e$ is the ellipsoidal concentration parameter. JS02 then
  provide a separate model for $c_e$. The independence of $R_0$ and
  $\delta_e^c$ can now be seen, since although both parameters depend
  on $c_e$, $R_0$ also depends on $R_e$, and $\delta_e^c$ also depends
  on $\Delta_e$. Thus owing to the independence of $R_0$ and
  $\delta_e^c$, for a given triaxial halo with axis ratios $a/c$ and
  $b/c$, we no longer have a simple method for determining the cut-off
  radius for the haloes from the initial mass, since the volume
  averaged over-density is not preserved. Instead, we must determine
  $R_{\vir}$ by numerically inverting the relation
  \be M=4\pi\rhob
  \frac{ab}{c^2}R_0^3\delta_e^c\left[\log(1+y)-y/(1+y)\right] \ ; \ \ 
  y\equiv \frac{R_{\vir}}{R_0}\ .\ee

  Note that again the definition of halo mass adopted by JS02 is
  different to that which we have adopted, where the JS02 mass is
  defined to be 
  \be M_{\JS}=\frac{4}{3}\pi r_{\vir}^3\rhob \Delta_{\vir}(\Omega)\ ,\ee
  However, the mapping between the masses can be found through solving
  the relation
  \be \left(\frac{c_{\JS}}{c_{\ST}}\right)^3 \frac{\Delta_{\vir}}{200}
  \left[\frac{\log(1+c_{\ST})-c_{\ST}/(1+c_{\ST})}
    {\log(1+c_{\JS})-c_{\JS}/(1+c_{\JS})}\right]=1\ ,\ee
  where $c_{\JS}=R^{\JS}_{\vir}/R_0$.  The mapping between the two
  concentration parameters is then simply given by
  \be c_{\ST}=\left(\frac{\Delta_{\vir}(\Omega)}{200}
  \frac{M_{\ST}}{M_{\JS}} \right)^{1/3}c_{\JS}\ .\ee

  JS02 also provided a convenient fitting formulae for the
  probability density functions for the axis ratios $a/c$ and $a/b$
  and these we use throughout. As a final note, JS02 advocate that
  $c_e$ be drawn from a log-normal distribution with mean concentration
  $\bar{c}_e$.  Since the effects of a stochastic concentration
  parameter have already been investigated in the halo model by
  \citet{CoorayHu2001}, we do not consider this here, and simply take
  $c_e=\bar{c}_e$.

\end{itemize}


\subsection{Correlation function for halo alignments}\label{ssec:corralign}

Following the discussion of intrinsic halo alignments in Section
\ref{ssec:align}, we now describe the toy model for $C(\mu)$. We seek
a function, defined on the range $\mu = [-1,+1]$, that is symmetric
about $\mu=0$ that obeys the integral constraint, $\int d\mu
C(\mu)=0$, and that gives rise to a uniform distribution of $\mu$ when
alignments are weak.  The simplest functional form that will allow us
to model these effects is to let $C(\mu)$ be linear in $|\mu|$.
Hence,
\be C(\mu|M_1,M_2) = \kappa(M_1,M_2)\left[|\mu|- \frac{1}{2}\right] 
\ ; \ -1\le\mu\le1
\label{eq:linIA}\ee
where $\kappa$ describes the slope of the relation and is in general a
function of $M_1$ and $M_2$.  

We now argue for how the mass dependence of $\kappa$ may arise and
suggest a possible form for its scaling.  Within the framework of the
extended Press-Schechter model \citep{Bondetal1991,LaceyCole1993}, one
expects that at any given redshift, the highest mass haloes are more
aligned than the lower-mass haloes. This can be understood on the
grounds that since lower mass haloes form at higher redshifts they
have more time to virialize and hence, particle orbits will become
isotropic. On the other hand, higher mass haloes form at lower
redshifts and therefore their shapes should still maintain some memory
of the tidal fields that they collapsed within and this leads to a
spin/shape alignment in nearby haloes.

As a consistency argument to this picture, we remark that in the work
of JS02 it is found that low-mass haloes are more spherical than
high-mass haloes and therefore, since no particular alignment can be
assigned to these objects, we should expect the correlation function
to be smaller than for the higher-mass objects. A further piece of
evidence supporting this view, comes from the numerical simulation
work of \citet{HattonNinin2001}, who observed a positive
mass-dependent trend in the correlation of angular momentum
vectors of dark matter haloes with local large-scale structure.  While
this is not exactly what we are considering here, it is at least
suggestive of a similar effect. 

Finally, we will make the further assumption that the dominant
alignment effect occurs between haloes of the same mass, and we will
therefore introduce a Dirac delta function $\delta^D(M_1-M_2)$ into
equation (\ref{eq:linIA}).  One may argue for this as follows: it is
more likely that haloes that collapse out of the same tidal field and
at the \emph{same time} will have their spin/shape vectors aligned.
However, the veracity of this assumption has yet to be established.
Nevertheless, the benefit of this is that the mass dependence of
$\kappa$ can be reduced to a simple $two-parameter$ function of $M$,
which means that $C(\mu)$ is also specified by just two parameters.
 
With these ideas in mind, we take $\kappa$ to be a power-law in
$\nu\equiv\delta_{\rm crit}/\sigma(M)$, where $\delta_{\rm crit}$ is
the linear theory collapse density from the spherical model and
$\sigma^2(M)$ is the linear theory variance of mass
fluctuations. Hence,
\be \kappa(\nu)=\kappa_0\, \nu^{\epsilon}\ ,\ee
where $\epsilon>0$ for high mass haloes to be more aligned and where
$\kappa_0$ sets the normalization for the $M_*$ haloes.  Also, as
$\nu\rightarrow 0$, $\kappa \rightarrow 0$, and we get, as required,
that $C(\mu)$ vanishes and the distribution of $\mu$ is uniform over
$[-1,+1]$. Given the lack of detailed information concerning halo
alignment correlations, it is difficult to assign values for
$\epsilon$ and $\kappa_0$ with any certainty. In what follows, we
therefore simply explore effects for the parameter choices:
$\epsilon=1.0,0.5,0.0$ and $\kappa_0=2.0,\,1.0,\,0.5$.

Before continuing, we note that more recent work by \citet{Jing2002}
and \citet{Hopkinsetal2004} has shown that there is indeed a mass
dependence in the alignment of haloes and that it is broadly in
agreement with the picture that we have described above. Furthermore,
the studies by \citet{HattonNinin2001} and \citet{Hopkinsetal2004}
have shown that the alignment effect decreases with radius as a
power-law. We would now like to emphasize that this is exactly what
our alignment model predicts. This is to be understood by the fact
that the alignment correlation function at zero spatial lag, which is
given by equation (\ref{eq:linIA}), is modulated by the cluster
correlation function (see equation \ref{eq:alignprob}). Thus according
to our model, the general alignment correlation function should simply
be a scaled multiple of the cluster correlation function. We intend to
fully explore this elsewhere.


\begin{figure*}
  \centerline{ \includegraphics[width=7.45cm]{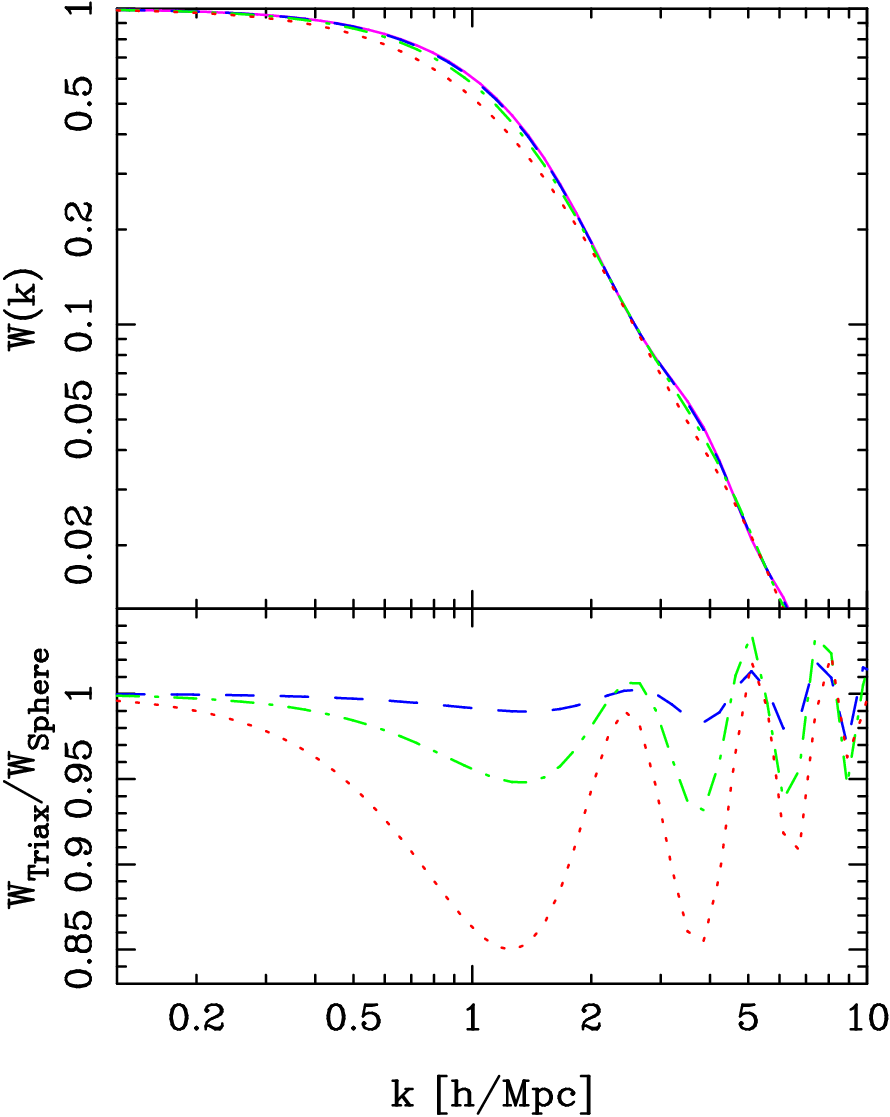}
    \includegraphics[width=7.45cm]{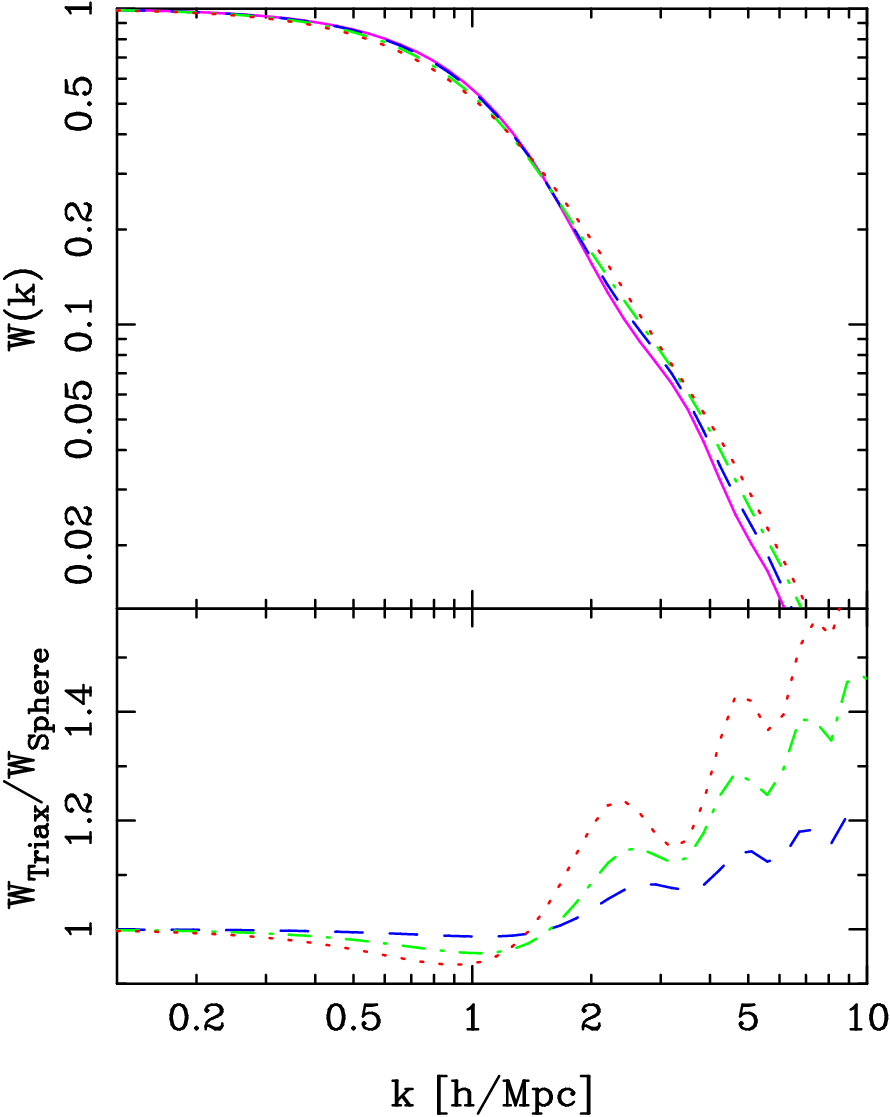}}
\caption{\small{Variation of the window functions with increasing
    prolaticity. Left panels show effects for the continuity density
    profile model and right show the density profile model of JS02. In
    the top panels, the solid line represents spherical haloes, and
    the dash, dot-dash and dotted lines represent haloes with
    $a/c=b/c=0.8,\, 0.6,\, 0.4$ respectively. The bottom panels show
    the ratio of the window functions with the equivalent spherical
    haloes from each model and line styles have been preserved. In
    all cases the halo mass was set to be $M=1.0\times10^{15}
    h^{-1}M\odot$.}
\label{fig:TriaxWindow}}
\end{figure*}


\begin{figure*}
  \centerline{\includegraphics[width=7.3cm]{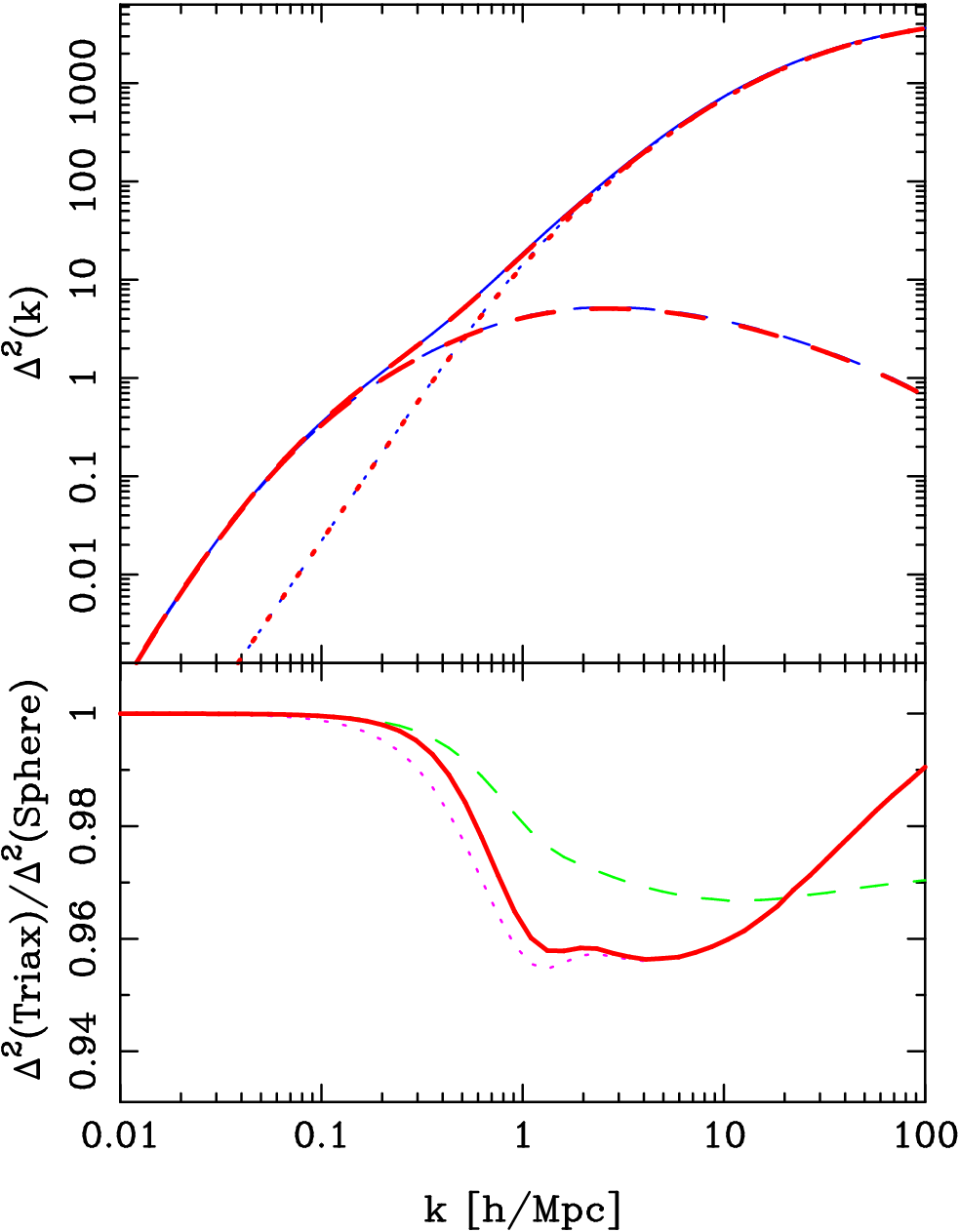}
    \includegraphics[width=7.3cm]{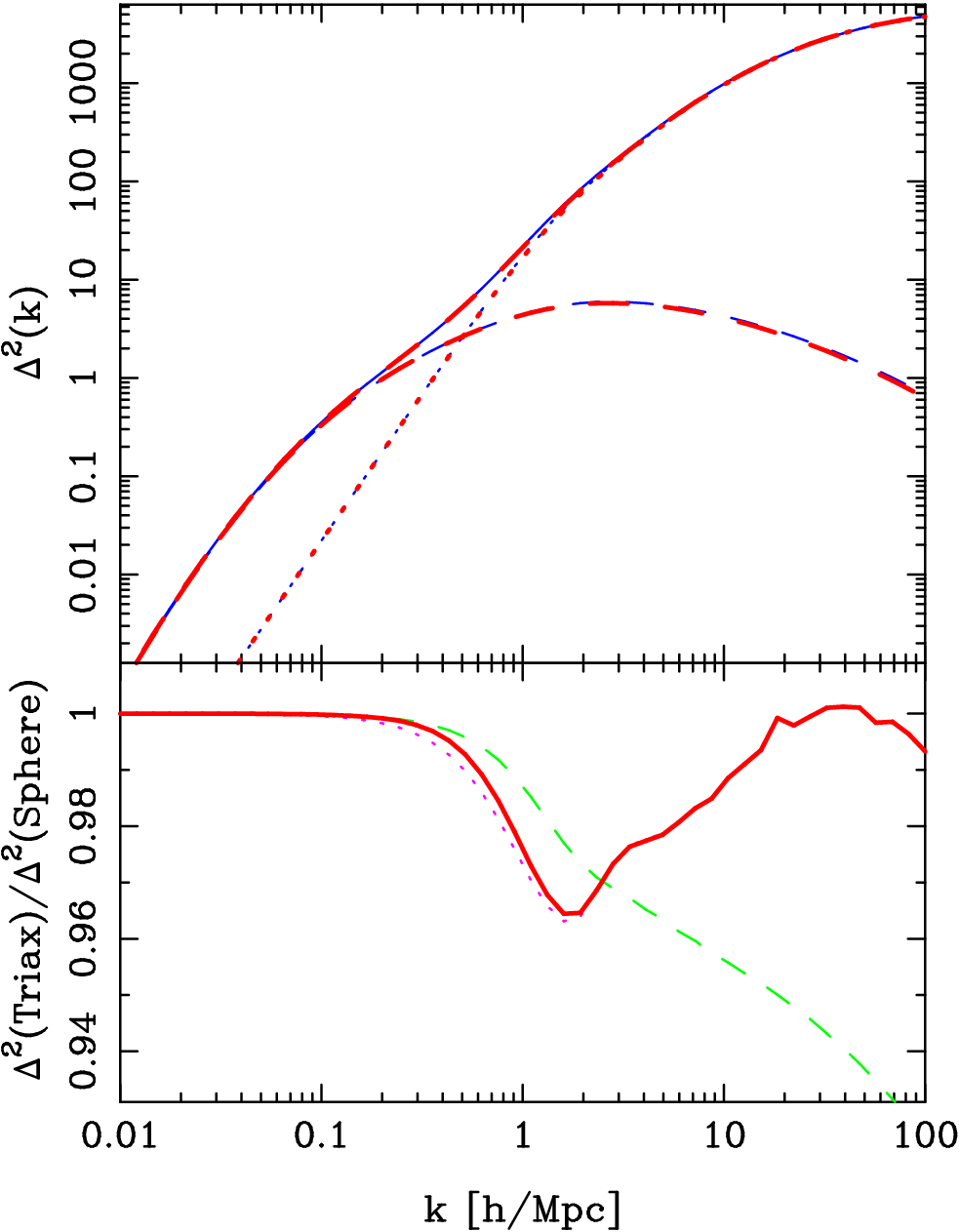}}
\caption{\small{Dimensionless power spectrum in the triaxial halo
    model. Left panels show effects for the continuity profile model
    and right show the JS02 profile model. Top panels show: the total
    power spectrum (thick solid line); which is the sum of $P^{\rm
    1H}$ (dotted lines) and $P^{\rm 2H}_{\rm NA}$ (dash lines); the
    total power from the spherical halo model (thin solid lines).
    Bottom panels show the ratios: total triaxial halo power to total
    spherical halo power (solid lines); triaxial 1-Halo to spherical
    1-Halo (dotted lines); triaxial 2-Halo to spherical 2-Halo (dash
    lines). }
\label{fig:TriaxPower}}
\end{figure*}


\begin{figure*}
  \centerline{\includegraphics[width=9cm]{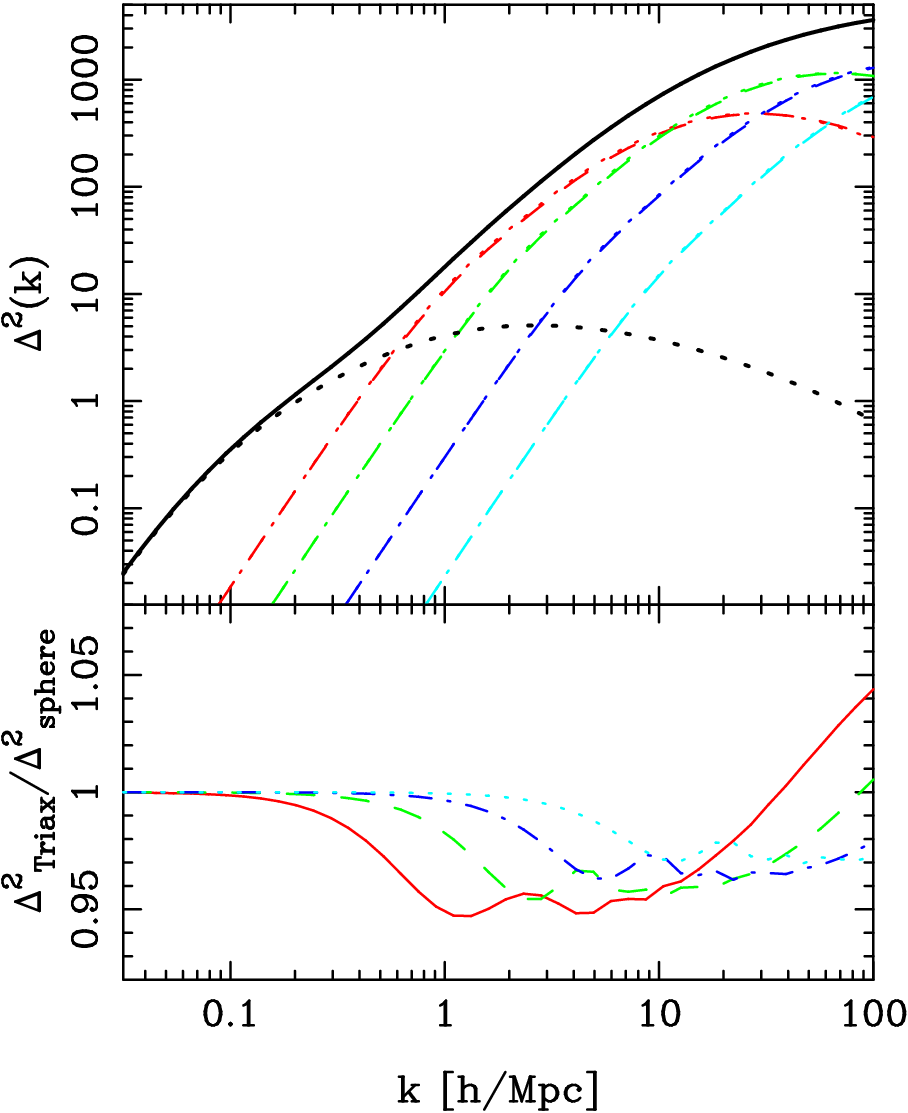}
    \includegraphics[width=9cm]{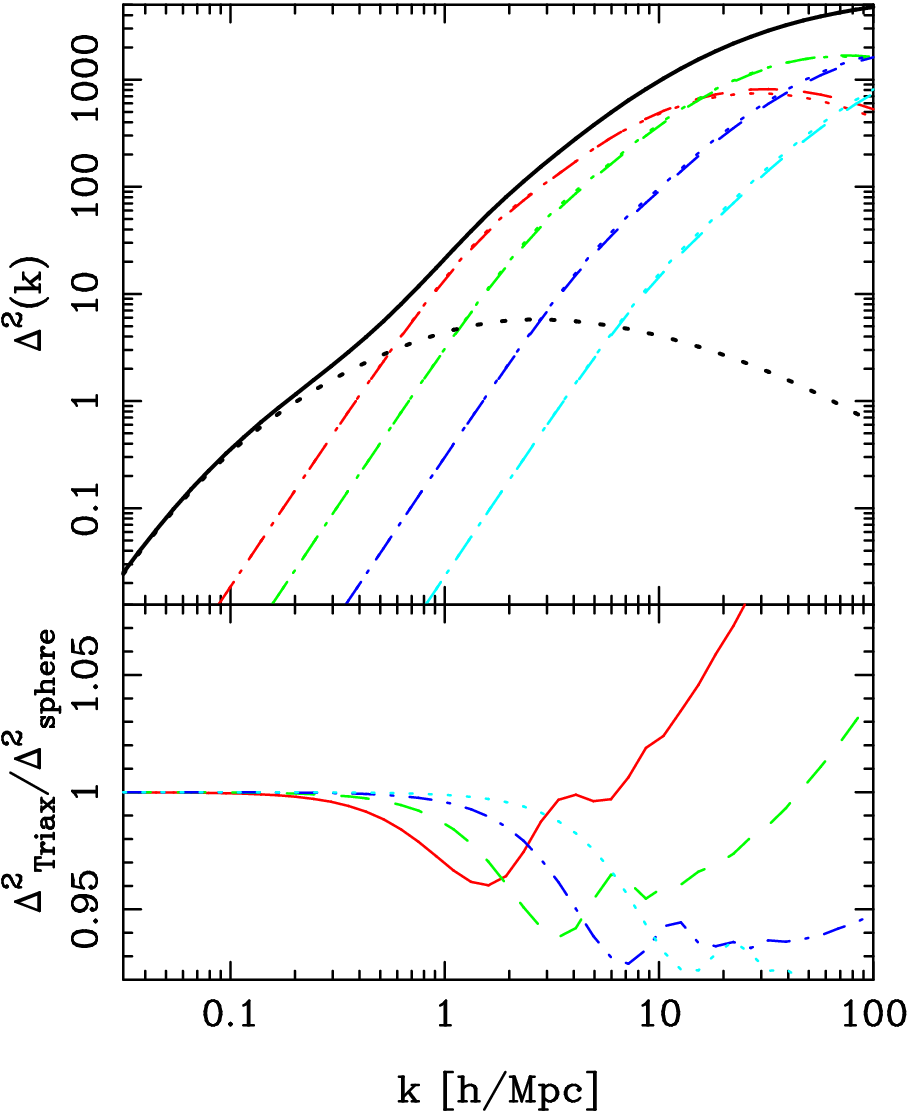}}
\caption{\small{Contribution to the triaxial power spectrum from
    different mass ranges. Left panels show the continuity profile and
    the right show the JS02 profile model. Top panels, the dash-line
    and the dotted lines, going from left to right, show the
    contribution to the 1-Halo term for the mass ranges: $M> 10^{14}
    h^{-1}M_{\odot}$; $10^{14}h^{-1}M_{\odot}\ge M>
    10^{13}h^{-1}M_{\odot}$; $10^{13}h^{-1}M_{\odot}\ge M>
    10^{12}h^{-1}M_{\odot}$; $10^{12}h^{-1}M_{\odot}\ge M>
    10^{11}h^{-1}M_{\odot}$. The thin dotted lines are the power
    spectrum split by mass for the equivalent spherical haloes. The
    thick dotted line is the 2-Halo term. Bottom panels show ratios
    of triaxial halo power to spherical for the same mass ranges as
    above, where we have adopted different and obvious line styles.}
\label{fig:TriaxPowerByMass}}
\end{figure*}


\begin{figure*}
  \centerline{\includegraphics[width=9cm]{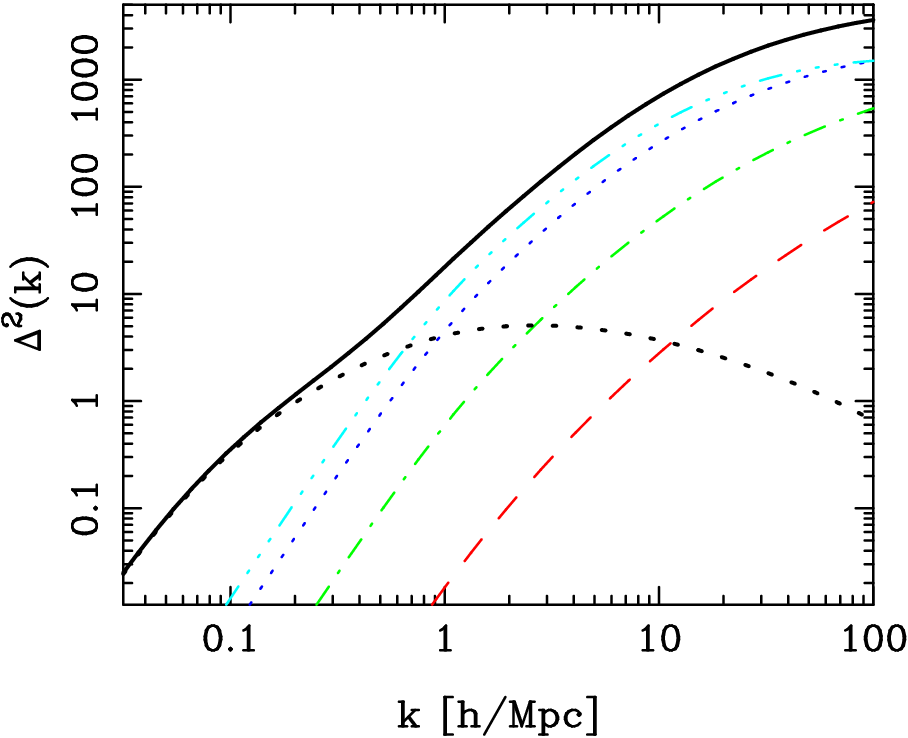}
    \includegraphics[width=9cm]{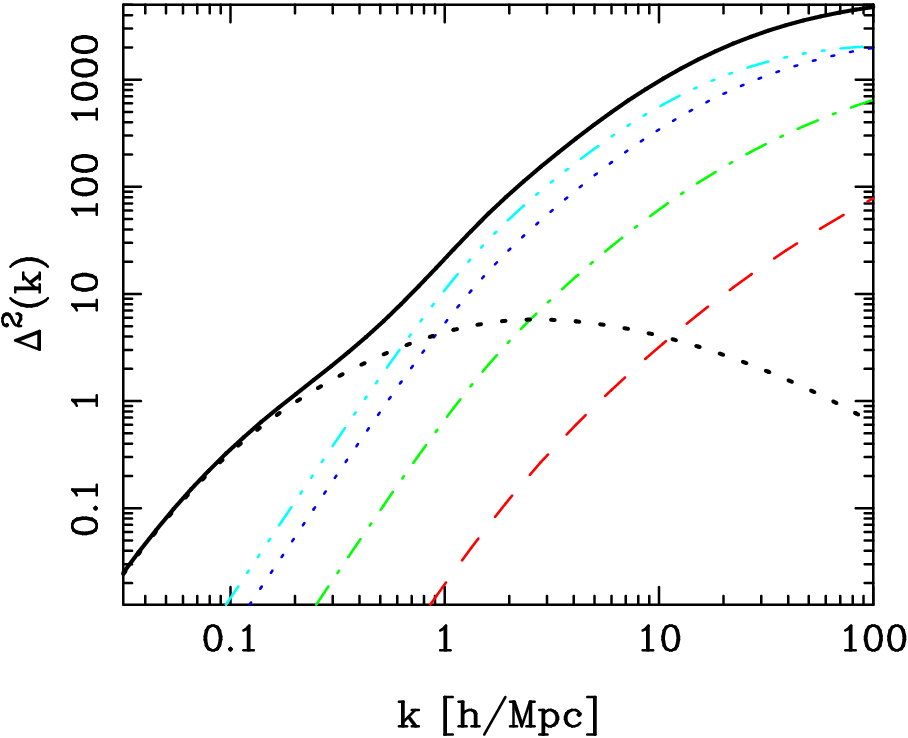}}
\caption{\small{Similar to Fig. \ref{fig:TriaxPowerByMass}, but this
    time showing the break down of the power spectrum with axis ratio
    $a/c$. In both panels, we show, going up from the dash line
    to the triple-dot dash line, the contribution to the 1-Halo term
    coming from haloes with axis ratios: $1.0\ge a/c>0.85$; $0.85\ge
    a/c>0.7 $; $0.7\ge a/c > 0.55 $; $0.55 \ge a/c >0.3 $. Again the
    left panel is the continuity model and the right is the JS02
    model.}
\label{fig:TriaxPowerByAxis}}
\end{figure*}


\begin{figure*}
  \centerline{\includegraphics[width=9cm]{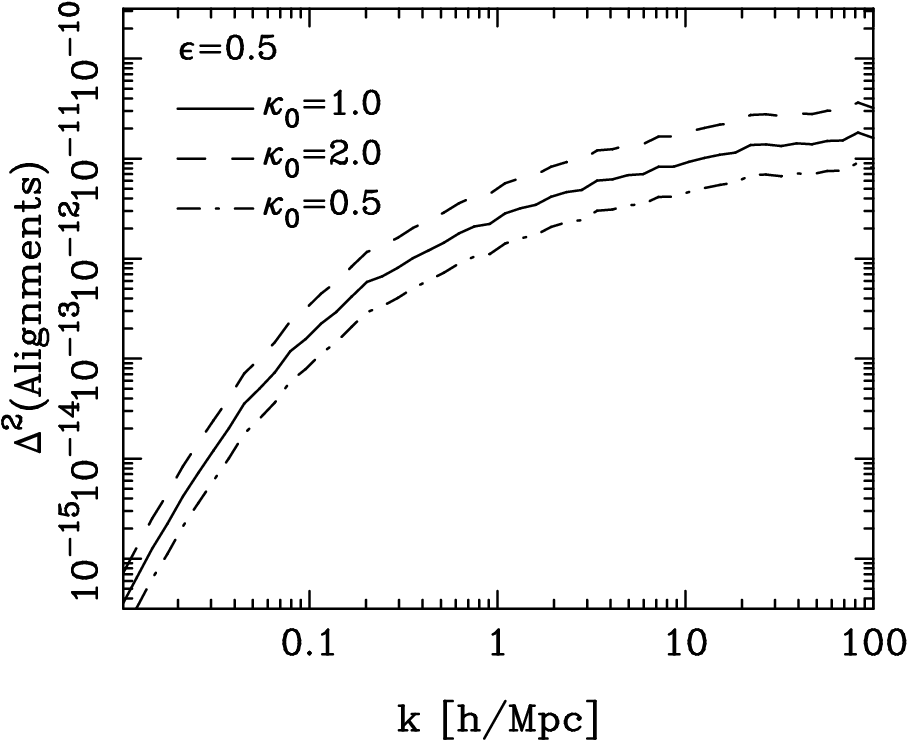}
    \includegraphics[width=9cm]{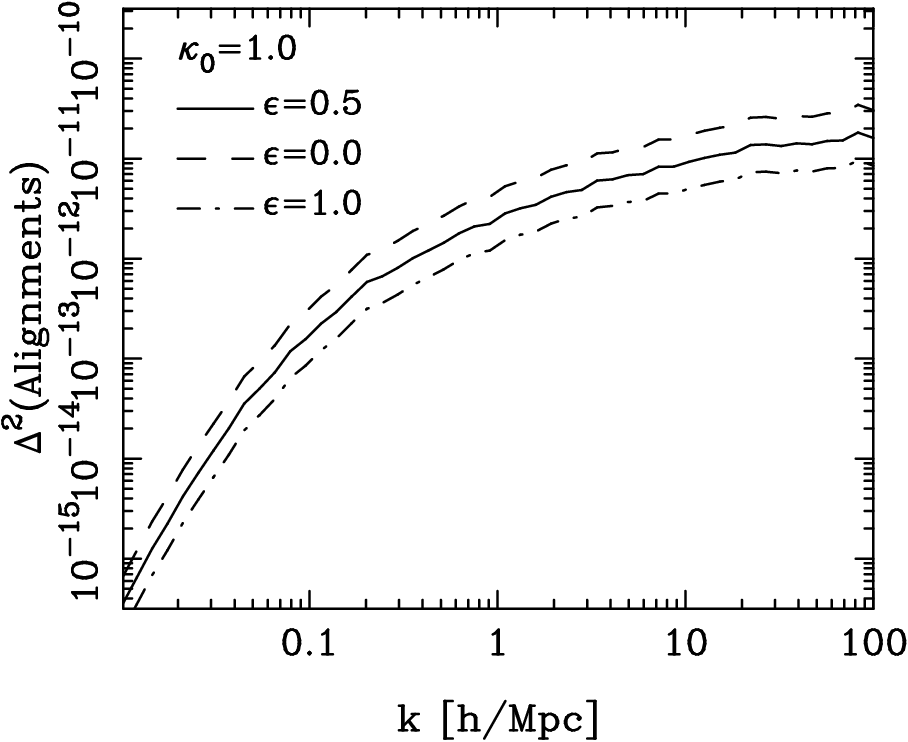}}
  \caption{\small{Contribution to the power spectrum from intrinsic
      alignments of the triaxial haloes, using the linear alignment
      model of equation (\ref{eq:linIA}). Left panel shows the
      variation with alignment normalization $\kappa_0$, and the right
      panel shows the variation with the power-law index $\epsilon$.}
    \label{fig:IntrinsicPower}}
\end{figure*}


\section{Results}\label{sec:results}

\subsection{Window functions \& power spectra}

In this Section, we present results for the numerical integration of
the 1-Halo term's window function, given by equation
(\ref{windowfunction}), and then the full power spectra given by
equations (\ref{power1H}), (\ref{power2Hnoalign}) and
(\ref{power2Halignments}). However, before proceeding, we feel that it
is necessary to mention the numerical algorithms we use to solve the
equations.

Although our analytical efforts have reduced the dimensionality of the
various power spectrum integrals dramatically, from 12- to 6-D for
$P^{\rm 1H}$ and $P^{\rm 2H}_{\rm NA}$ and from 18- to 12-D for
$P^{\rm 2H}_{\rm IA}$, the numerical integration of these equations
using serial quadratures is extremely inefficient and intensely
demanding on cpu time. We therefore recommend the use of an efficient
multi-dimensional integrator, such as the Korobov-Conroy algorithm
\citep{Korobov1963,Conroy1967} or the Sag-Szekeres algorithm
\citep{SagSzekeres1964}, both of which allow fast and accurate
evaluations of these expressions (taking $\sim$10 minutes to evaluate
equation (\ref{power2Halignments}) on a standard 2.8Ghz 32-bit Intel
processor with heavy optimization afforded through the Intel
compiler, and where the number of evaluation points in the
Sag-Szekeres algorithm was set to 100 million).  Both of the above
algorithms have implementations vended through the Numerical
Algorithms Group, listed as routines: {\ttfamily d01fdf} and
{\ttfamily d01gcf}.

Figure \ref{fig:TriaxWindow} shows how, for the two different density
profile models considered, the window functions, specified by equation
(\ref{windowfunction}), change as haloes become more prolate. From
inspecting the top panels of these figures, it is clear that the
overall effect is small. We therefore take the ratio of the window
functions with those for the equivalent spherical model to inspect the
effect more closely, where by equivalent spherical model we mean the
window function for a particular triaxial profile model with
$a/c=b/c=1$. For the continuity profile model (bottom left panel), we
find that as the prolaticity increases, the ratio gets smaller. This
relative suppression in the window function is maximal on scales of the
order the virial radius of the halo, being at most $\sim$15\% for the
extreme, $a/c=0.4$, prolate objects.  Considering the haloes with JS02
profiles (right panels), we find that the effect is first seen as a
small suppression on large scales and then as an amplification on small
scales.  We ascribe this amplification to the fact that in the JS02
model $\delta_e^c$ increases as haloes become more ellipsoidal.

Figure \ref{fig:TriaxPower} shows the total power spectrum in the
triaxial halo model with no intrinsic alignments, obtained by the sum
of equations (\ref{power1H}) and (\ref{power2Hnoalign}). Note that
here we plot the dimensionless power spectrum: $\Delta^2(k)=4\pi
k^3P(k)/(2\pi)^3$ \citep{Peacock1999}.  Considering the triaxial
haloes with the continuity profile model (left panels), again the
effect is weak. Looking at the ratio with the spherical case, we show
that it is maximal on scales of the order the virial radius of
clusters and is manifest as a suppression of power at the $\sim5\%$
level.  The oscillatory features that were seen in the window function
have been washed out by averaging over the halo mass and axis ratio
distributions.  Considering the resultant power spectrum from the JS02
model (right panels), we find a similar overall effect as for the
continuity model, with a suppression of the order $\sim5\%$ on scales
$k\sim2 h {\rm Mpc}^{-1}$. The oscillatory features seen in the
window function again have been smoothed out and the small scale
amplification is no longer apparent.  For the case where haloes are not
intrinsically aligned, we find differences between the spherical
2-Halo term and the triaxial of the order a few percent for both
models. However, these effects occur on scales where the 2-Halo power
is sub-dominant to the 1-Halo power and therefore can be neglected.

Figure \ref{fig:TriaxPowerByMass}, shows the contributions to the
1-Halo term from different mass ranges. For the continuity profile
model (left panel), we find for each mass bin considered, that the
power is suppressed relative to the spherical case, and that the
effect is strongest for the high mass haloes, being at the level
$\sim6\%$. There is also an small amplification of power on small
scales. For the JS02 profile model (right panel), we find that there
is a characteristic de-amplification, followed by a strong
amplification of power as one goes from large to small
scales. Interestingly, the large-scale suppression of power is
stronger for the lower mass haloes, being of the order $\sim8\%$ for
the $M\sim5.0\times10^{11}h^{-1}M_{\odot}$ objects on scales $k\sim10h
{\rm Mpc}^{-1}$. However, the small scale amplification effect is
strongest for the high mass haloes, boosting the power to around
$\sim10\%$ for the $M\sim 10^{15}h^{-1}M_{\odot}$ on similar scales.
Thus one sees that the largest effects of triaxiality are most
apparent in the highest mass haloes. This result follows in accordance
with the hierarchical picture of structure formation: high mass haloes
form at late times in the Universe and thus particle orbits have
insufficient time to circularize and hence structures are more likely
to be triaxial than spherical. This effect is built into the JS02
probability density function through a mass dependent scaling of the
stochastic variable $a/c$, which operates so that higher mass haloes
are more likely to be triaxial than the lower mass haloes (see
equations 16 and 17 in JS02). This indicates that if one is purly
intersted in the clustering properties of extreme mass objects, then
halo triaxiality plays a more significant role.

Figure \ref{fig:TriaxPowerByAxis} shows the contribution to the 1-Halo
term from different ranges of the axis ratio $a/c$. Somewhat
interestingly, in both cases, for $k<10\, h {\rm Mpc}^{-1}$ the power
spectrum is entirely dominated by the most triaxial haloes, $a/c<0.7$,
and that the spherical haloes contribute very little to the overall
power. This observed effect: highly triaxial haloes contributing the
most on large scales and spherical haloes accounting for more on small
scales, can be understood entirely from the mass dependent scaling of
the distribuion functions for $a/c$.

Figure \ref{fig:IntrinsicPower} shows the contributions to the power
spectrum from the 2-Halo intrinsic alignment term. We find that, for
the linear correlation model specified by equation (\ref{eq:linIA}),
the intrinsic alignments contribution to the power spectrum is,
$\sim10$ orders of magnitude smaller than the 2-Halo term with no
alignmnets. As will become apparent from the following sub-section,
this result is actually insensitive to our choices for $\kappa_0$ and
$\epsilon$. Considering the variation of the clustering with
$\kappa_0$ (left panel), we find that increasing/decreasing $\kappa_0$
simply uniformally increases/decreases $P^{\rm 2H}_{\rm IA}$.  Next,
considering the variation of the power with the power-law index
$\epsilon$ (right panel), we find that as $\epsilon$
increases/decreases the alignment power spectrum decreases/increases,
and that the variations are very similar to those with
$\kappa_0$. This may mean that $\epsilon$ and $\kappa_0$ are not
independent parameters.


\subsection{Maximally alligned haloes}

As a corollory to this section we discuss the limiting case of a
maximally aligned halo distribution, that is, one in which the
orientation vector of every halo is perfectly alligned with the
others. Clearly, this scenario is unphysical, however it does allow us
to place firm constraints on the largest possible contribution that
alignments may provide.  In order to achieve this we simply set the
correlation function to be
\be C(\mu)=\delta^D(\mu\pm 1)-1\ \ ; \hspace{0.4cm}-1\le\mu\le1\ .\ee
In this limit we find that the function ${\mathcal D}$ in the
intrinsic alignment term given by (equation \ref{power2Halignments})
becomes
\ba {\mathcal D}(\theta_k, \phi_k) & = & 4\pi \int d\psi\, 
\,\sbess0\left[kR'f(\theta_k'',\phi_k'')\right] \nonumber \\
& &\ \  - \int d\alpha d(\cos\beta) d\gamma
\sbess0\left[kR' f(\theta_k',\phi_k')\right], \ea
where ${\bf k}'' = {\mathcal R}(0,0,\psi){\bf k} $ and ${\bf k}' =
{\mathcal R}(\alpha,\beta,\gamma){\bf k} $, and where we have used the
fact that for $\mu = \pm 1$, we may combine rotations over $\alpha$
and $\gamma$ into a new rotation over $\psi = \alpha - \gamma$.

\begin{figure}
  \includegraphics[width=8.4cm]{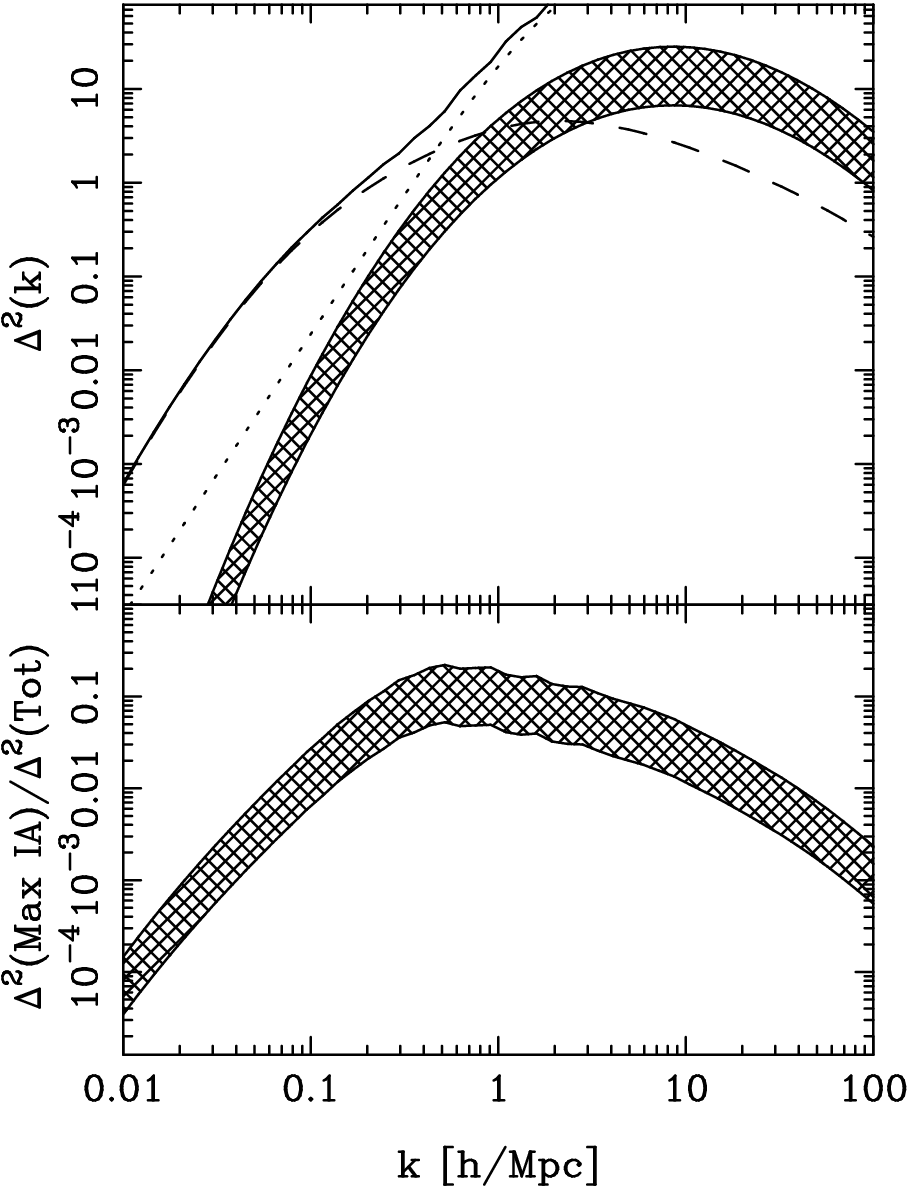}
  \caption{\small{Maximal contribution to the power spectrum from
      intrinsic halo alignments. The top panel shows: the maximum
      alignments (cross hatched region defines the result $\pm$ the
      \emph{rms} error); the 2-Halo term with no alignments (dash line);
      the 1-Halo term (dotted line); and the total contribution from all
      terms (thin solid line). The bottom panel shows the ratio of the
      maximum to the total power.  The triaxial density profile model
      was the continuity model.}
    \label{fig:Maximal}}
\end{figure}

Figure \ref{fig:Maximal} shows the maximum possible contribution to
the power spectrum from the intrinsic alignments of dark matter haloes
that have the continuity density profile. We see that for this case,
the alignment contribution has significantly increased by roughly
$\sim8-10$ orders of magnitude. However, relative to the total power
(bottom panel), it is still a fairly small quantity, being at most an
effect of the order $\sim10\%$ on scales $k\sim 1 \,h\,$Mpc$^{-1}$.
Given that this alignment correlation function is physically
unrealistic, we may now conclude that the true contribution to the
power spectrum from alignments must be $<10\%$.


\section{Discussion \& Conclusions}\label{sec:discussion}

In this paper we have developed a formalism for calculating the
clustering statistics of a distribution of intrinsically aligned,
triaxial, dark matter haloes: the triaxial halo model. This formalism
facilitates the exploration of the importance of halo alignments and
shapes on the density clustering statistics. 

As a direct application of the formalism, we have considered the
lowest order clustering statistic of interest, that is the power
spectrum. We found, as usual, that the general result separates into a
term that describes the clustering between haloes (2-Halo) and a term
that describes the clustering within a halo (1-Halo). However, with
the inclusion of intrinsic halo alignments we found that the 2-Halo
term itself separates into two terms: the first described the
halo-halo clustering without alignments; the second described the
clustering due to halo alignments.  We have derived compact analytic
forms for these general relations, in all cases dramatically reducing
the dimensionality of the integral equations. These were then solved
numerically using two different multi-dimensional integrators.  We
considered two different density profile models. The first allowed us
to explore pure shape effects on the clustering, the second was the
more realistic profile model of JS02, which modified the halo central
density based on halo shape. In both cases, we found that the effects
on the power spectrum were, as expected, small, being at most
$\sim5\%$ for $k\sim1-10 h\,$Mpc$^{-1}$. However, when considerd by
mass we found a more significant effect for the high mass objects,
with a suppression of the order $\sim6\%$ for the continuity model and
an $\sim5\%$ suppression of power on large scales followed by a
$\sim10-15\%$ amplification of power on small-scales for the JS02
model. The effects of halo triaxiality will be important to account
for when interpreting precision measurements of cluster correlation
functions on small scales. We have also found that the 1-Halo power is
dominated by haloes with $a/c<0.7$.

We have explored the impact of halo alignments on the power spectrum.
In order to achieve this it was necessary to develop a toy-model for
the correlation function of the semi-major axis direction vectors of
haloes. This model was constructed so that high mass haloes were more
likely to be alligned than lower mass haloes. We further made the
assumption that only haloes of the same mass are alligned.  Although
questionable, this assumption provided a means to consider the effects
of a mass dependent alignmnet correlation function.  For this model,
we found that halo alignment contribution to the power spectrum was,
surprisingly, $\sim10$ orders of magnitude smaller than the 2-Halo
term without alignments. Modifying the normalization parameters for
the $C(\mu)$ does not significantly increase the alignment
power. Thus, if our alignment model is correct halo alignments are
completely unimportant for density clustering statistics.

We then constructed the maximal alignment correlation function. This
allowed us to explore the physically unrealistic case where all haloes
of all masses are completely alligned with one another. We found, for
the triaxial haloes with the continuity profile, that the maximum
alignment contribution to the power spectrum was $\sim10\%$ on scales
of the order $k\sim 1\,h\,$Mpc$^{-1}$ and significantly less on all
other scales. This lead us to the conclusion that the true alignment
contribution to the power must be $<10\%$.

We now add an important and necessary cautionary note. In order to
make accurate predictions for the clustering in the triaxial halo
model, we require an accurate model for the statistical properties of
the triaxial dark matter haloes.  Clearly, the first density profile
model that we considered, the continuity model, was constructed simply
as a toy-model, and as such the predictions should not be expected to
match reality. In the second model, the JS02 model, the profiles were
constructed to match results from numerical simulations. However, we
have discovered some aspects of this model that should be clarified
before it can be used with confidence to make accurate predictions for
the clustering statistics.

Firstly, consider the special case of a halo that is actually
spherical and apply the JS02 formalism to it, having obtained the two
independent normalization parameters, we ask the question: At what
radius does the average overdensity reach that at which we defined the
halo? The answer is not the same as the virial radius that we defined
from the mass.  Secondly, the ellipsoidal concentration parameter
$c_e\equiv R_e/R_0$ depends on the axis ratio $a/c$, and not the
second ratio $b/c$. This leads to the following problem: When
averaging over the axis ratio distributions, we found that whilst a
haloes shape and characteristc density may change dramatically through
changes in $b/c$, the concentration paramter remains fixed. This
results in inconsistant density structures for the haloes, since we
found that haloes of a given mass and $a/c$ could be less and more
dense than spherical haloes, depending on the value of $b/c$. In
reality it is more likely that haloes that are triaxial are either
less dense, or of equivalent density, or more dense than spherical
haloes, but not all. Whilst the model of JS02 is ground breaking in
many ways, we feel in light of these problems that some aspects should
be re-visited.

For the 1-Halo term we have found that modelling the density structure
of haloes with triaxial ellipsoids produces an effect of the order
$5-20\%$, relative to equivalently defined spherical haloes. For
observational programs that hope to measure the small scale power
spectrum to these levels of accuracy or better, one must therefore
account for halo triaxiality when interpreting data with halo models.

We mentioned earlier that if one takes halo concentration to be a
stochastic variable in the spherical halo model, then this too
increases the small scale power \citep{CoorayHu2001}. The effect is
thus degenerate with triaxiality for the JS02 haloes.  If the main
reason for the stochasticity of the halo concentration could be
attributed to the incorrect assumption that haloes are spherical, then
one might understand the degeneracy. However, JS02 found that there is
a comparable scatter in the concentraion parameter for the ellipsoidal
haloes as there is for the spherical. This they suggest is due to
differences in the merger histories of the haloes and not asphericity.

A further effect on the small scale clustering that we have not yet
mentioned is that of halo subtructures. The halo model formalism was
extended to include this by \citet{ShethJain2003}. Effects on the
density power spectrum were then explored by \citet{Dolneyetal2004}.
They found that, in general, the small-scale clustering becomes
dominated by substructures below a certain scale. The scale depends on
the sub-structure mass function and density distribuiton. Again, the
effect appears to be degenerate with the effects of triaxiality.

In this paper we have considered only the lowest order clustering
statistics, the 2-pt auto-correalation function and the power
spectrum.  However, the overall goal of this work is to explore how
the shapes of dark matter haloes and their intrinsic alignments
influence the hierarchy of correlation functions. Since the 2-pt
clustering statistics are the least sensative to the shapes of
structures, it is not surprising that the degeneracies, noted above,
have been found. However, it is expected that these will be broken
through consideration of the higher order statistics, such as the
bispectrum, and we will focus on this in a subsequent paper.


\section*{acknowledgements}
We thank Martin White for a useful discussion prior to the outset of
this work. We also thank John Peacock for useful discussions during
this work. RES acknowledges the PPARC for postdoctoral research
assistantships. PIRW thanks the PPARC and the University of Bonn for
research assistantships.


\setlength{\bibhang}{2.0em}


\appendix

\section{Rotation matrix}\label{app:rotation}
Owing to there being several equivalent ways to define the Euler
angles for the rotation matrix ${\mathcal R}(\alpha,\beta,\gamma)$, we
make explicit the definition that we use throughout. The matrix for
the $z-y'-z''$ rotation is given by:

\be {\mathcal R}(\alpha,\beta,\gamma)\equiv \left(
\begin{array}{ccc}

\c\,\beta\, \c\,\alpha \c\,\gamma & \c\,\beta\, \s\,\alpha\, \c\,\gamma & 
-\s\,\beta\, \c\,\gamma\\
-\s\,\alpha\, \s\,\gamma & +\c\,\alpha\, \s\,\gamma & \\
-\c\,\beta\, \c\,\alpha\, \s\,\gamma & -\c\,\beta\, \s\,\alpha 
\,\s\,\gamma & \s\,\beta\, \s\,\gamma \\
-\s\,\alpha\, \c\,\gamma & + \c\,\alpha\,  \c\,\gamma & \\
\s\,\beta\, \c\,\alpha & \s\,\beta\, \s\,\alpha & \c\,\beta \\
\end{array}\right)
\ ,\ee
where we have adopted the short hand notation $\c=\cos$ and $\s=\sin$.


\section{The mass function and halo biasing}\label{app:massfunction}

We model the mass function of dark matter haloes using the model of
ST, since the predictions are in excellent agreement with the halo
abundances measured directly from numerical simulations
\citep[ST][]{Jenkinsetal2001,Reedetal2003}.  The ST mass function
is
\be \frac{dn(M)}{d\log M}=\frac{\rhob}{M}\;\nu f(\nu)
\left|\frac{d\ln\nu}{d\ln M}\right|\ ;\ \
\nu\equiv\dcol/\sigma(M)\ ,\ee
where
\be \nu f(\nu)= 2A \left(1+\frac{1}{\nu'^{2q}}\right)\left(\frac{\nu'^{2}}{2\pi}\right)
\exp\left(-\frac{\nu'^2}{2}\right) \ , \ee
and where $\nu'=\sqrt{a}\nu$, $a=0.707$, $q=0.3$ and $A\approx0.322$.
\citet{ShethMoTormen2000} argued that the success of the ST formula
was due to the fact that the ellipsoidal model for halo collapse was
more realistic, as opposed to the spherical.

We will also require a model for the halo biasing. Here we again look
to the work of ST, who found

\be b(M)=1+\frac{a\nu^2-1}{\dcol}+\frac{2p/\dcol}{1+(a\nu^2)^p}\ ;\ \
\ \nu\equiv \dcol/\sigma\ . \ee

\end{document}